\documentclass[11pt]{article}
\usepackage{graphicx}        
\usepackage{latexsym}
\usepackage{amsmath}
\usepackage{amsfonts}
\usepackage{amssymb}
\advance\textheight by \topskip
\begin{document}
\baselineskip 0.8cm
\thispagestyle{empty}
\begin{flushright}
KUCP0220\\
UTAP-422\\
October 3, 2002\\ 
\end{flushright}
\vskip 2 cm
\begin{center}
{\LARGE \bf Cosmology with Radion and Bulk Scalar Field} \\
{\LARGE \bf in Two Branes Model}
\vskip 1.7cm

{\bf Shinpei Kobayashi}
\footnote{E-mail: shinpei@phys.h.kyoto-u.ac.jp}
{\bf and}\ 
{\bf Kazuya Koyama}
\footnote{E-mail: kazuya@utap.phys.s.u-tokyo.ac.jp}\

\vskip 1.5mm

\vskip 2cm
 $^1$ Graduate School of Human and Environment Studies, Kyoto
University, Kyoto  606-8501, Japan \\
 $^2$ Department of Physics, The University of Tokyo, Tokyo 113-0033, 
     Japan 
\end{center}

\newpage

\vskip 1cm
{\centerline{\large\bf Abstract}}
\begin{quotation}
\vskip -0.4cm
We investigate cosmological evolutions of the 
bulk scalar field $\phi(t)$ and the radion $d(t)$ in five-dimensional 
dilatonic two branes model. The bulk potential for the scalar field is 
taken as the exponential function $V_{bulk} \propto \exp(-2 \sqrt{2} b\phi)$,
where $b$ is the parameter of the theory. 
This model includes Randall-Sundrum model (with $b=0$) and five-dimensional 
Ho\v{r}ava-Witten theory (with $b=1$). We consider matter on both 
branes and arbitrary potentials on the branes and in the bulk. These matter and 
potentials induce the cosmological expansion of the brane as well as the 
time evolution of the bulk scalar field and the radion. Starting with full 
five-dimensional equations, we derive four-dimensional 
effective equations which govern the low-energy dynamics of brane worlds. 
A correspondent five-dimensional geometry is also obtained. The effective 
four-dimensional theory on a positive tension brane is described by bi-scalar 
tensor theory.  If the radion is stabilized, the effective theory becomes 
Brans-Dicke (BD) theory with BD parameter $1/2 b^2$. On the other hand, 
if the scalar field is stabilized, the effective theory becomes scalar-tensor theory 
with BD parameter $\frac{3}{2(3b^2+1)}\frac{\varphi(t)}{1-\varphi(t)}$ where $\varphi$ is 
the BD field defined by radion $d(t)$. If we do not introduce the stabilization
mechanism for these moduli fields, the acceptable late
time cosmology can be realized only if the dilaton coupling $b$ is small
($b^2 < 1.6 \times 10^{-4}$) and the negative tension brane is sufficiently
away from the positive tension brane. We also construct several models for
inflationary brane worlds driven by potentials on the brane and in the bulk. 

\end{quotation}

\newpage

\section{Introduction}

String Theory and/or M-Theory are thought to be the most promising 
candidate for ``The Theory of Everything''. One of the interesting 
and embarrassing features of such theories is that they are consistently 
formulated in more than four-dimensions. On the other hand, at least 
at low energy scales, our spacetime seems to be four-dimensional. 
Hence we need some compactification mechanisms of extra dimensions 
in order to retrieve our four-dimensional spacetime at low energy
scales. 
 
Of all compactification mechanisms of the extra dimensions, 
Ho\v{r}ava-Witten (HW) theory is a fascinating one\cite{HW1}\cite{HW2}. 
Ho\v{r}ava and Witten showed that the 
eleven-dimensional limit of M-theory compactified on $S^1/{\bf Z}_2$ 
orbifold can be identified as the strongly coupled ten-dimensional 
$E_8 \times E_8$ heterotic string theory with two orbifold fixed planes.  
Witten also showed that there exists a consistent compactification of this 
theory where the universe appears as a five-dimensional spacetime 
with two boundary branes \cite{Witten}. Then the five dimensional 
effective action for this theory was derived \cite{Lukas1} and 
the cosmological evolution of the branes has been studied 
\cite{Lukas2}-\cite{Reall}. \footnote{ 
Note that the time evolution of  their cosmological 
domain wall solution is driven by the bulk scalar field, 
not by the matter on the brane. So their solution is a ``vacuum 
cosmological solution'', which is different from the cosmological 
solution in our usual sense. }\ 
In this model, the bulk scalar field is necessarily associated with the 
compactification. 

The bulk scalar field 
also plays an important role in phenomenological models for brane worlds.
For example, the bulk inflaton models 
\cite{KKS}-\cite{SHS}, the self-tuning models for cosmological constant 
\cite{uzawa}-\cite{BCG} and the dilatonic thick domain wall models \cite{KKS2}
-\cite{CEHS} have been studied as the extension of
Randall-Sundrum (RS) models \cite{RS1}, \cite{RS2}. Cosmological models 
with bulk scalars have been actively considered \cite{LM},\cite{MB}. 
In the context of RS model, stabilization mechanism of the 
distance between two branes using bulk scalars has been 
investigated \cite{GW1}-\cite{CF}. 

As well as the bulk scalar field, there is another important 
scalar degree of freedom in two branes model, that is,  a radion 
which describes the inter-brane distance. The radion plays a 
crucial role in gravitational interactions on the brane. 
For example, in RS model, the radion acts as a Brans-Dicke (BD) field
and the observational constraints on BD parameter gives strong constraint
on the model. In RS model, the dynamics of the radion has been eagerly 
studied \cite{CG}-\cite{K}. 

The purpose of this paper is to investigate the cosmological consequences 
of the bulk scalar field and the radion. We consider a model where the 
bulk potential for the bulk scalar field $\phi$ is given by an exponential 
function $V_{bulk} \propto \exp (-2 \sqrt{2} b \phi)$. This model includes  
HW theory (with $b=1$) and RS model (with $b=0$). We will consider matter on both 
branes and arbitrary potentials on the branes and in the bulk. These matter and 
potentials induce the cosmological expansion of the brane as well as the 
time evolution of the bulk scalar field and the radion. It is well known that 
in many higher-dimensional theories, these "moduli fields" are potentially
dangerous because time evolution of moduli fields leads to the time variation
of fundamental constants. Then it is important to know how serious is the 
moduli problem in this model. We also consider the phenomenological 
implication of the potentials mainly focusing on the inflationary scenario.

In order to know the cosmological consequences of the 
bulk scalar field and radion, we should derive the effective
equations which govern the cosmological dynamics of these  
fields. One approach is to construct an 
4D Effective action by the integration of the action with respect to 
the extra dimension \cite{chiba}-\cite{BBDR1}.  
A ``moduli space'' approximation is an example for such an approach.
In this approximation,  the constant moduli parameters 
of the static solution is replaced by time-dependent functions and integration 
over the extra dimensions gives four-dimensional effective action. 
But, it is known that the moduli space
approximation may not always be consistent \cite{KKLT}-\cite{T}
and the integration over extra-dimension must be treated with special caution 
\cite{Mukohyama}.

Hence, in this paper, we will derive four-dimensional effective 
equations for the radion and the bulk scalar field starting with full 
five-dimensional equations. In order to do so, we use the low-energy expansion 
scheme. The low-energy means that the energy density of the matter 
on the brane 
is much lower than the tension of the brane. This relation holds 
at late times in cosmological evolution. Under such an approximation, 
we can derive the effective four-dimensional equations for 
scale factor, radion and bulk scalar field. We can also determine
the correspondent five-dimensional geometry.

The organization of this paper as follows. In Sec. II, we review the setup 
of our model and introduce the low-energy approximation. Using 
this approximation, we give the background solutions.  
In Sec. III, we solve the first-order equations and derive  
the solutions for full five-dimensional metric at this order.
Using these solutions, we show that 
the equations for scale factor on the brane, bulk scalar
and radion can be derived. 
Sec. IV is the main part of this paper. Effective equations 
which govern the low energy dynamics of the two branes system
are presented. The comparison with scalar-tensor theory 
is performed and the observational constraints are obtained.  
Sec. V, several inflationary scenarios are discussed.
Sec. VI is devoted to conclusions.

\section{Setup}

\subsection{Action and Equations of Motion}

We consider the bulk action of the form
\begin{equation}
S_{bulk} = \frac{1}{2\kappa^2}\int_{M_5} d^5 x \sqrt{-g_5}
\left(R-\partial_M \phi \partial^M \phi + V_{bulk}(\phi) \right),
\end{equation} 
where $\kappa^2$ denotes the five-dimensional gravitational constant 
and $\sigma$ denotes the tension of the 3-brane.
The bulk potential is taken as 
\begin{equation}
V_{bulk}(\phi) = -\left(b^2 -\frac{2}{3}\right)
e^{-2\sqrt{2}b \phi} \sigma^2 +V(\phi). 
\end{equation} 
Here we split $V_{bulk}$ into an explicit, exponential potential 
for $\phi$ and an arbitrary potential $V(\phi)$ according to 
Ref.\cite{LOW}. 
The parameter $b$ is determined by the theory.  
If we choose $b=0$ and $V(\phi)=0$, we retrieve the 
action for the Randall-Sundrum model and if $b=1$ and $V(\phi)=0$, 
we retrieve the action for Ho\v{r}ava-Witten model, 
which were derived in \cite{Lukas1}.

We consider a five-dimensional spacetime with structure 
$M_5 = S^1/{\bf Z}_2 \times M_4$ where $S^1/{\bf Z}_2$ is an 
orbifold and $M_4$ is a smooth manifold. We use the coordinate $y$ to 
parameterize the orbifold and the range of $y$ is chosen as 
$-r \le y \le r$. Because this spacetime has ${\bf Z}_2$-symmetry (i.e. we  
identify $y$ with $-y$), there are two endpoints in this spacetime.  
These two endpoints are called orbifold fixed planes, 
which are 3-branes. A positive tension brane is set at $y=0$ 
and a negative tension brane is set at $y=r$. We take the actions for them as  
\begin{eqnarray}
S_{brane} &=& S_+ + S_- \nonumber \\
S_+ &=&  \int_{M_4^{(1)}} d^4 x 
                        \sqrt{-g_4} U_{brane} 
+\int_{M_4^{(1)}} d^4 x \sqrt{-g_4} {\cal L}_m, \\ \nonumber
S_- &=&  \int_{M_4^{(2)}} d^4 x
                        \sqrt{-g_4} \tilde{U}_{brane} 
+\int_{M_4^{(2)}} d^4 x \sqrt{-g_4} \tilde{{\cal L}}_m, 
\end{eqnarray}
where the potentials are given by
\begin{eqnarray}
U_{brane} &=& -\frac{\sqrt{2}}{\kappa^2} e^{-\sqrt{2} b \phi} \sigma
-U(\phi), \nonumber\\
\tilde{U}_{brane} &=& \frac{\sqrt{2}}{\kappa^2} e^{-\sqrt{2} b \phi} \sigma
-\tilde{U}(\phi). 
\end{eqnarray}
The action for the positive tension brane and the negative tension brane 
is denoted by $S_{+}$ and $S_{-}$, respectively.  
$U(\phi)$ and $\tilde{U}(\phi)$ are arbitrary potentials. 
${\cal L}_m$ and $\tilde{{\cal L}}_m$ 
are the Lagrangian density of the matter on each brane. 
We assume that they do not couple to the bulk scalar, that is, 
$\delta {\cal L}_m/\delta \phi 
= \delta {\tilde{\cal L}}_m/\delta \phi =0$.  
It is true that in HW model, there are  
field strengths of $E_8$ and $E_6$ gauge field which couple to 
the bulk scalar field, but in this paper, we replace them 
with the energy density of perfect fluids  
\footnote{
It is also interesting to consider the coupling of the gauge 
field and the scalar field. It is left for the 
future work. 
}\ . 

Using the action, we derive the Einstein equation 
\begin{equation}
G_{\mbox{ }N}^{M} = 
\kappa^2 (T_{bulk \mbox{ }N}^{\mbox{ }\mbox{ }\mbox{ }M} 
+ T_{brane \mbox{ }N}^{\mbox{ }\mbox{ }\mbox{ }M}). 
\end{equation}
The energy-momentum tensor can be divided into two parts;
\begin{eqnarray}
T_{bulk \mbox{ }N}^{\mbox{ }\mbox{ }\mbox{ }M} &=& \frac{1}{\kappa^2}
\left\{
\nabla^M \phi \nabla_N \phi 
-\frac{1}{2}\delta_{\mbox{ }N}^{M}
\left[
\nabla^M \phi \nabla_M \phi 
+\left(
b^2 -\frac{2}{3}
\right)
 e^{-2\sqrt{2}b\phi}\sigma^2
-V(\phi)
\right]
\right\}, \nonumber \\
T_{brane \mbox{ }N}^{\mbox{ }\mbox{ }\mbox{ }M} &=& 
\frac{\sqrt{-g_4}}{\sqrt{-g_5}}
\Bigg\{
\left[
-\frac{\sqrt{2}}{\kappa^2}e^{-\sqrt{2}b\phi}\ 
diag(0,\sigma,\sigma,\sigma,\sigma)
+diag(0,-\rho-U,p-U,p-U,p-U)
\right]\delta (y) \nonumber \\
\mbox{ }\hspace{1cm} & & +\left[
\frac{\sqrt{2}}{\kappa^2}e^{-\sqrt{2}b\phi} \ 
diag(0,\sigma,\sigma,\sigma,\sigma)
+diag(0,-\tilde{\rho}-\tilde{U},\tilde{p}-\tilde{U},
\tilde{p}-\tilde{U},\tilde{p}-\tilde{U})
\right]\delta (y-r)
\Bigg\}.
\end{eqnarray}
The energy density and the pressure of the perfect 
fluid on the positive tension brane are denoted by 
$\rho$ and $p$ and those on the negative tension brane
are denoted by $\tilde{\rho}$ and $\tilde{p}$. 
The equation of motion for the scalar field is given by 
\begin{multline}
\nabla^2 \phi 
+ \sqrt{2}b\left(b^2-\frac{2}{3}\right)
e^{-2\sqrt{2} b \phi} \sigma^2 +\frac{1}{2} \frac{d V}{d \phi} \nonumber \\
=\frac{\sqrt{-g_4}}{\sqrt{-g_5}}
 \left[
  \left(
   -2be^{-\sqrt{2} b \phi} \sigma 
   +\kappa^2 \frac{d U}{d \phi}
  \right) \delta (y) 
  +\left(
   2be^{-\sqrt{2} b \phi} \sigma 
   +\kappa^2 \frac{d \tilde{U}}{d \phi}
  \right) \delta (y-r) 
 \right]. 
\end{multline}

Now we take the five-dimensional metric as 
\begin{equation}
ds^2 = e^{2\gamma(y, t)}dy^2 -e^{2\beta(y,t )}dt^2 
            +e^{2\alpha(y, t)}\delta_{ij}dx^i dx^j. 
\end{equation}
Einstein equations and the scalar field equation in terms of 
$\alpha, \beta, \gamma$ and $\phi$ are presented in 
Appendix A-1. We expand the functions around $y=0$ and $y=r$ as    
\begin{eqnarray}
\alpha(y,t) &=& \alpha^{(0)}(t)+ \alpha^{(1)}(t)|y|
               +\frac{1}{2}\alpha^{(2)}(t)y^2 + \cdots 
\label{eq:around_0}, \\
\alpha(y,t) &=& \tilde{\alpha}^{(0)}(t)
                     + \tilde{\alpha}^{(1)}(t)|r-y|
                 + \frac{1}{2}\tilde{\alpha}^{(2)}(t)(r-y)^2 + \cdots. 
\label{eq:around_r}  
\end{eqnarray}
From above expansions, we can read that 
$\alpha^{\prime\prime}$ gives $2\delta(y)-2\delta(y-r)$. Then 
from Einstein equations 
$\alpha^{(1)}$ and $\tilde{\alpha}^{(1)}$ 
can be written in terms of $\rho$, $p$, $U$ and $\tilde{\rho}$, 
$\tilde{p}$, $\tilde{U}$ respectively. Then we can get the junction conditions; 
\begin{eqnarray}
\alpha^{(1)}(t) &=& 
-\frac{1}{6}
 \left[
  \sqrt{2} \sigma e^{-\sqrt{2}b \phi} 
  + \kappa^2 (\rho + U )
 \right]
 e^{\gamma} \bigg|_{y=0},  \\
\beta^{(1)}(t) &=& 
-\frac{1}{6}
 \left[
  \sqrt{2}\sigma e^{-\sqrt{2}b \phi} 
  - \kappa^2 (3p +2\rho - U )
 \right]
 e^{\gamma} \bigg|_{y=0},  \\
\phi^{(1)}(t) &=& 
-\left[
  b \sigma e^{-\sqrt{2}b \phi} 
  - \frac{1}{2} \kappa^2 \frac{dU}{d\phi}
 \right]
 e^{\gamma} \bigg|_{y=0},  \\
\tilde{\alpha}^{(1)}(t) &=& 
-\frac{1}{6}
 \left[
  \sqrt{2} \sigma e^{-\sqrt{2}b \phi} 
  - \kappa^2 (\tilde{\rho} + \tilde{U} )
 \right]
 e^{\gamma} \bigg|_{y=r},  \\
\tilde{\beta}^{(1)}(t) &=& 
-\frac{1}{6}
 \left[
  \sqrt{2} \sigma e^{-\sqrt{2}b \phi} 
  + \kappa^2 (3 \tilde{p} +2 \tilde{\rho} - \tilde{U} )
 \right]
 e^{\gamma} \bigg|_{y=r},  \\
\tilde{\phi}^{(1)}(t) &=& 
-\left[
  b \sigma e^{-\sqrt{2}b \phi} 
  + \frac{1}{2} \kappa^2 \frac{d \tilde{U}}{d\phi}
 \right]
 e^{\gamma} \bigg|_{y=r}.
\end{eqnarray}

\subsection{Low Energy Expansion Scheme}

In this subsection, we formulate the low energy expansion scheme to 
solve the equations \footnote{
Generalizations to covariant form of this iteration scheme 
are performed in \cite{Kanno1}, \cite{Kanno2}}\ . 
At late times, the energy density and 
the pressure of the perfect fluid is much smaller than the 
tension of the brane. Hence we can impose the following conditions,    
\begin{equation}
\sigma \gg \kappa^2 \rho,\ \kappa^2 p,\ \sigma^{-1} V,\ 
\kappa^2 U,\ \kappa^2 \tilde{U}.
\end{equation}
Then, we can define a small parameter of the system as follows,  
\begin{equation}
\varepsilon \equiv \frac{\kappa^2 \rho}{\sigma}. 
\end{equation}
Thus for $\varepsilon \ll 1$, we can solve five-dimensional Einstein 
equations perturbatively by expanding a metric and scalar field as  
\begin{eqnarray}
\alpha(y,t) &=& \alpha_{(0)}(y,t)+ \alpha_{(1)}(y,t) 
+ \alpha_{(2)}(y,t) +\cdots ,
\label{eq:expansion}
\end{eqnarray}
where $\alpha_{(i)}$ is the function of the order ${\cal O} (\varepsilon^i)$.
Note that the lower indices $\alpha_{(i)}$ are different from the upper 
indices $\alpha^{(i)}$ which have been defined in eqs.(\ref{eq:around_0}),
(\ref{eq:around_r}). 
We expand $\beta(y,t),\ \gamma(y,t),$ and $\phi(y,t)$ in a similar
way. The important point is that $t$-derivative of the lower
order solution is the same order as $y$-derivative of the higher
order solution;
\begin{equation}
\ddot{\alpha}_{(i-1)} \sim \alpha^{''}_{(i)}.
\end{equation}
This is because the derivative 
with respect to $y$
is of the order $\alpha' \sim \sigma$ for $\varepsilon \ll 1$ 
from the junction conditions.
On the other hand, from the Friedmann equation, we expect 
that the derivative with respect to $t$ is of the order 
$\dot{\alpha}^2 \sim \kappa^2 \sigma \rho$. Then the derivative
with respect to $t$ is suppressed by the factor $\varepsilon^{1/2}$
compared with the derivative with respect to $y$;
\begin{equation}
\left( \frac{\dot{\alpha}}{\alpha'} \right)^2  \sim \varepsilon.
\end{equation}
Hence, the leading order equation of the order ${\cal O}(\epsilon^0)$ 
contains only the $y$-derivative of the 0-th order solution
(i.e., $\alpha_{(0)}^{\prime\prime},\ 
\beta_{(0)}^{\prime\prime}, \cdots$),  
and 1st order equations are consisted of $t$-derivatives of 0-th 
order functions and $y$-derivatives of 1st order functions 
(i.e., $\ddot{\alpha}_{(0)}, \ \ddot{\beta}_{(0)}, \cdots, \ 
 \alpha_{(1)}^{\prime\prime},\ \ \beta_{(1)}^{\prime\prime}, 
\cdots$) as shown in Appendix \ref{APP1}. 
At each order, Einstein equations give 
ordinary differential equations for ($\alpha_{(i)},\ \beta_{(i)},
\cdots$) with respect to $y$. Then it is possible to solve the 
equations analytically.

\subsection{Solution at Zeroth Order} 

The concrete form of the 0-th order equations are given in 
Appendix A-2 and the solutions at the 0-th order are given 
as follows, 
\begin{eqnarray}
\alpha_{(0)}(y,t) &=& \frac{2}{3\Delta +8}\ln H(y,t) + \hat{\alpha}(t), \\
\beta_{(0)}(y,t) &=& \frac{2}{3\Delta +8}\ln H(y,t),  \\
\gamma_{(0)}(y,t) &=& \ln d(t) + \sqrt{2}b\hat{\phi}(t), \\
\phi_{(0)}(y,t) &=& \frac{1}{\sqrt{2}b}\ln H(y,t) + \hat{\phi}(t), 
\end{eqnarray}
where, 
\begin{equation}
H(y,t) \equiv 1 - \frac{\sqrt{2}(3\Delta +8)}{12}\ d(t)\ \sigma \ |y|, 
\end{equation}
and 
\begin{equation}
\Delta \equiv 4\left(b^2-\frac{2}{3} \right). 
\end{equation}
Here $\hat{\alpha}(t)$ can be identified with the scale factor 
on a positive tension brane and $\hat{\phi}(t)$ describes the behavior of the 
scalar field. $d(t)$ describes the time evolution of inter-brane distance. 
There does not exist $\hat{\beta}(t)$ because $\beta_{(0)}$ corresponds 
to the lapse function, so we can always gauge away $\hat{\beta}(t)$ 
by the redefinition of the time variable $t$. 
These zeroth-order solutions are quasi-static and their time evolutions 
will be determined by the first-order solutions.  
If we take $\hat{\alpha}(t)=\hat{\phi}(t)=0$ and $d(t)=1$, we can 
retrieve the vacuum solution obtained in \cite{cvetic}-\cite{CLP}. 

The above solutions are valid except for RS model with 
$b =0 (\Delta=-8/3)$. For RS model, the 0-th order solutions
are given by 
\begin{eqnarray}
\alpha_{(0)}(y,t) &=& -d(t) k |y|  + \hat{\alpha}(t), \\
\beta_{(0)}(y,t) &=& -d(t) k|y|,  \\
\gamma_{(0)}(y,t) &=& \ln d(t), \\
\phi_{(0)}(y,t) &=& \hat{\phi}(t),
\end{eqnarray} 
where $k=\sqrt{2} \sigma/6$ is the curvature scale of the bulk Anti-
de Sitter spacetime. In the following, we assume $b \neq 0$ and RS model is 
treated separately in Appendix C. 

\section{Solutions for bulk metric and scalar field}

\subsection{Solutions at First Order}

\subsubsection{Equations}

The concrete form of the 1-th order equations are given in 
Appendix A-3. 
Substituting the solutions at 0-th order into the equations at 1st
order, we get 
\begin{multline}
-\frac{\sqrt{2}}{6}\frac{\sigma d}{H}
(3\alpha_{(1)}^{\prime}+\beta_{(1)}^{\prime}
-\sqrt{2}b\phi_{(1)}^{\prime})
+\frac{1}{3}\left(b^2-\frac{2}{3}\right)
\frac{\sigma^2 d^2}{H^2}(\gamma_{(1)}-\sqrt{2}b\phi_{(1)}) \\
=d^2 e^{2\sqrt{2}b\hat{\phi}}
\Bigg\{
H^{-\frac{1}{3b^2}}
\Bigg[
\ddot{\hat{\alpha}} +2 \dot{\hat{\alpha}}^2 
+ \frac{1}{6}\dot{\hat{\phi}}^2 
-\frac{\sqrt{2}}{6}\frac{\sigma}{H}y
(\ddot{d}+3\dot{d}\dot{\hat{\alpha}}+\sqrt{2}b\dot{d}\dot{\hat{\phi}}) \\
+\frac{1-3b^2}{18}\frac{\sigma^2}{H^2}\dot{d}^2 y^2
\Bigg]
+\frac{1}{6}V \Bigg\},
\label{eq:yy}
\end{multline}
\begin{multline}
\alpha_{(1)}^{\prime\prime}
-\frac{\sqrt{2}}{6}\frac{\sigma d}{H}
(4\alpha_{(1)}^{\prime}-\gamma_{(1)}^{\prime}+\sqrt{2}b\phi_{(1)}^{\prime})
+\frac{1}{3}\left(b^2-\frac{2}{3}\right)
\frac{\sigma^2 d^2}{H^2}(\gamma_{(1)}-\sqrt{2}b\phi_{(1)}) \\
=d^2 e^{2\sqrt{2}b\hat{\phi}}
\Bigg\{ 
H^{-\frac{1}{3b^2}}
\Bigg[
\dot{\hat{\alpha}}^2 + \sqrt{2}\dot{\hat{\alpha}}\dot{\hat{\phi}} 
- \frac{1}{6}\dot{\hat{\phi}}^2 +\frac{\dot{d}}{d}\dot{\hat{\alpha}}
-\frac{\sqrt{2}}{6}\frac{\sigma}{H}y
\left(\frac{\dot{d}^2}{d}+2\dot{d}\dot{\hat{\alpha}} \right) \\
+\frac{1-3b^2}{18}\frac{\sigma^2}{H^2}\dot{d}^2 y^2
\Bigg]
+\frac{1}{6}V
\Bigg\},
\end{multline}
\begin{multline}
2\alpha_{(1)}^{\prime\prime}+\beta_{(1)}^{\prime\prime} 
-\frac{\sqrt{2}}{6}\frac{\sigma d}{H}
(8\alpha_{(1)}^{\prime}+4\beta_{(1)}^{\prime}-3\gamma_{(1)}^{\prime}
+3\sqrt{2}b\phi_{(1)}^{\prime}) \\
+\left(b^2-\frac{2}{3}\right)\frac{\sigma^2 d^2}{H^2}
(\gamma_{(1)}-\sqrt{2}b\phi_{(1)}) \\
=d^2 e^{2\sqrt{2}b\hat{\phi}}
\Bigg\{
H^{-\frac{1}{3b^2}}
\Bigg[
2\ddot{\hat{\alpha}}+2\sqrt{2}b\dot{\hat{\alpha}}\dot{\hat{\phi}}
+3\dot{\hat{\alpha}}^2+\sqrt{2}b\ddot{\hat{\phi}}
+\left(2b^2 +\frac{1}{2}\right)\dot{\hat{\phi}}^2 +2\frac{\dot{d}}{d}\dot{\hat{\alpha}}
+\frac{\ddot{d}}{d}+2\sqrt{2}b\dot{\hat{\phi}}\frac{\dot{d}}{d} \\
-\frac{\sqrt{2}}{6}\frac{\sigma}{H}y
\left(
2\ddot{d}+4\sqrt{2}b\dot{d}\dot{\hat{\phi}}
+4\dot{d}\dot{\hat{\alpha}}
+\frac{\dot{d}^2}{d}
\right) \\
+\frac{1-3b^2}{18}\frac{\sigma^2}{H^2}\dot{d}^2 y^2
\Bigg]
+\frac{1}{2} V
\Bigg\},
\end{multline}
\begin{multline}
\alpha_{(1)}^{\prime}
\left(
\sqrt{2}b\dot{\hat{\phi}}+\frac{\dot{d}}{d}
\right)
-\frac{1}{3}\phi_{(1)}^{\prime}
\left(
-b\frac{\sigma}{H}\dot{d}y+\dot{\hat{\phi}}
\right) \\
=
\dot{\alpha}_{(1)}^{\prime}
+\frac{\sqrt{2}}{6}\frac{\sigma d}{H}\left(
\dot{\gamma}_{(1)}-\sqrt{2}b\dot{\phi}_{(1)}
\right)
-\left(
-\frac{\sqrt{2}}{6}\frac{\sigma}{H}\dot{d}y+\dot{\hat{\alpha}}
\right)
\left(
\beta_{(1)}^{\prime}-\alpha_{(1)}^{\prime}
\right), 
\label{eq:yt-component}
\end{multline}
\begin{multline}
\phi_{(1)}^{\prime\prime}
-\frac{\sigma d}{H}
\left[b(3\alpha_{(1)}^{\prime}+\beta_{(1)}^{\prime}
-\gamma_{(1)}^{\prime}) +\frac{2\sqrt{2}}{3}\phi_{(1)}^{\prime} \right] 
+2\sqrt{2}b\left(b^2-\frac{2}{3}\right)\frac{\sigma^2 d^2}{H^2}
(\gamma_{(1)}-\sqrt{2}b\phi_{(1)}) \\
=d^2 e^{2\sqrt{2}b\hat{\phi}}
\Bigg( H^{-\frac{1}{3b^2}}
\Bigg\{
\ddot{\hat{\phi}}+3\dot{\hat{\alpha}}\dot{\hat{\phi}}
+\sqrt{2} b \dot{\hat{\phi}}^2 + \frac{\dot{d}}{d}\dot{\hat{\phi}} 
-\frac{\sigma}{H}y\left[b\left(
\ddot{d}+\sqrt{2}\dot{d}\dot{\hat{\phi}}
+3\dot{d}\dot{\hat{\alpha}}+\frac{\dot{d}^2}{d} \right)
+\frac{\sqrt{2}}{3}\dot{\hat{\phi}}\dot{d}
\right]  \\
+ \frac{\sqrt{2}}{3}b(1-3b^2)\frac{\sigma^2}{H^2}\dot{d}^2 y^2
\Bigg\}
-\frac{1}{2} \frac{dV}{d\phi}
\Bigg ).
\label{eq:phi}
\end{multline}
These equations can be regarded as the ordinary differential 
equations for $(\alpha_{(1)},\beta_{(1)},\gamma_{(1)},\phi_{(1)})$
where the right-hand sides of the equations act as sources. 
Then the solutions are given by homogeneous solutions and 
particular solutions which are determined by $\dot{\hat{\alpha}}, 
\dot{\hat{\phi}}, \dot{d}$ and $V(\phi_{(0)})$.

\subsubsection{Homogeneous Solutions}
The homogeneous solutions are given by
\begin{eqnarray}
\alpha_{(1)} &=& K_{\alpha}(t)H^{\frac{b^2-\frac{2}{3}}{b^2}}+F_{\alpha}(t)
                   + P(y,t), \\
\beta_{(1)} &=&  K_{\beta}(t)H^{\frac{b^2-\frac{2}{3}}{b^2}}+ F_{\beta}(t)+ P(y,t), \\
\gamma_{(1)} &=& K_{\gamma}(t)H^{\frac{b^2-\frac{2}{3}}{b^2}}
+  \sqrt{2}b F_{\phi}(t) + 6b^2 P(y,t), 
                     -3\sqrt{2}\frac{H}{\sigma d}\ P^{\prime}(y,t),\\
\phi_{(1)} &=& K_{\phi}(t)H^{\frac{b^2-\frac{2}{3}}{b^2}}+ F_{\phi}(t)+3\sqrt{2}b\ P(y,t). 
\end{eqnarray}
Here, $F_{\alpha},\  F_{\beta},\  F_{\phi}$ and $P$ are arbitrary 
functions and they do not affect the resultant effective equations. 
On the other hand, $K_{\alpha},\ K_{\beta},\  K_{\gamma},\ K_{\phi}$ 
carry the information of the bulk. These functions satisfy
two constraints; 
\begin{eqnarray}
3K_{\alpha}+K_{\beta}+K_{\gamma}-2\sqrt{2}b\ K_{\phi}&=& 0,\nonumber\\ 
\dot{K}_{\alpha}+(K_{\alpha}-K_{\beta})\dot{\hat{\alpha}} -
\left( \sqrt{2} b K_{\alpha}- \frac{1}{3} K_{\phi} \right)  
\dot{\hat{\phi}} 
&=& \frac{2}{3 \Delta}(\dot{K}_{\gamma}- \sqrt{2} b \dot{K}_{\phi}).
\label{eq:constraint}
\end{eqnarray}
These constraints are insufficient to determine the behavior of the 
homogeneous solutions. The homogeneous solutions will be determined
by the junctions conditions at the branes.

We should note that in the case of RS model 
($b=0$) without scalar field $K_{\phi}=0$, these constraints
are sufficient to obtain the concrete form of the homogeneous solutions.  
We obtain the equation for $K$ as 
\begin{equation}
\left( K_{\alpha} +\frac{1}{4} K_ {\gamma} \right)^{\cdot} 
+4 \dot{\hat{\alpha}} \left( K_{\alpha} + \frac{1}{4} K_{\gamma} \right) =0. 
\end{equation}
Thus we find 
\begin{equation}
K_{\alpha} +\frac{1}{4}K_{\gamma} = \frac{1}{8} C e^{-4\hat{\alpha}}. 
\end{equation}
where $C$ is the integration constant. It is well known that $C$
is related to the mass of the Black hole in the bulk and 
this term will induce ``dark radiation'' on the brane. 

\subsubsection{Particular Solutions}

From eqs.(\ref{eq:yy})-(\ref{eq:phi}), the particular solutions is given by 
\begin{eqnarray}
\alpha_{(1)} &=& f_{\alpha}(t)H^{2-(1/3b^2)}(y,t)
                 +g_{\alpha}(t)H^{1-(1/3b^2)}(y,t)y
                 +h_{\alpha}(t)H^{-1/3b^2}(y,t) y^2
                 +k_{\alpha}(y,t), 
\label{eq:particular_alpha} \\
\beta_{(1)} &=& f_{\beta}(t)H^{2-(1/3b^2)}(y,t)
                 +g_{\beta}(t)H^{1-(1/3b^2)}(y,t)y
                 +h_{\beta}(t)H^{-1/3b^2}(y,t) y^2 
                 +k_{\beta}(y,t), 
\label{eq:particular_beta} \\
\gamma_{(1)} &=& f_{\gamma}(t)H^{2-(1/3b^2)}(y,t)
                 +g_{\gamma}(t)H^{1-(1/3b^2)}(y,t)y
                 +h_{\gamma}(t)H^{-1/3b^2}(y,t) y^2
                 +\sqrt{2}bk_{\phi}(y,t), 
\label{eq:particular_gamma} \\
\phi_{(1)} &=& f_{\phi}(t)H^{2-(1/3b^2)}(y,t)
                 +g_{\phi}(t)H^{1-(1/3b^2)}(y,t)y
                 +h_{\phi}(t)H^{-1/3b^2}(y,t) y^2
                 +k_{\phi}(y,t). 
\label{eq:particular_phi}
\end{eqnarray}
The functions $(f_{\alpha}$, $g_{\alpha}$, 
$h_{\alpha}$, $f_{\beta}$, 
$g_{\beta}$, $h_{\beta}$, $f_{\gamma}$, $g_{\gamma}$, $h_{\gamma}$, 
$f_{\phi}$, 
$g_{\phi}$, $h_{\phi})$
are determined by $\hat{\alpha}(t), \ \hat{\phi}(t)$ and $d(t)$.
The solutions are presented in Appendix B. 
On the other hand, the solution $k(y,t)$ is determined by
bulk potential $V$ as 
\begin{eqnarray}
k_{\alpha}'(y,t) &=& \frac{1}{6} d^2 e^{2 \sqrt{2} b \hat{\phi}} 
H^{-2/3b^2} \int^y_0 dy H^{2/3 b^2} V,   \\ 
k_{\beta}'(y,t) &=&  -\frac{1}{2} d^2 e^{2 \sqrt{2} b \hat{\phi}}
H^{-2/3b^2} \int^y_0 dy H^{2/3 b^2} \left( 
\sqrt{2}b\frac{d V}{d \phi} + (1+2b^2) V \right) \nonumber  \\
& &\mbox{ }-\frac{\sqrt{2}}{2}\sigma^{-1}d e^{2\sqrt{2}b\hat{\phi}}H V, \\
k_{\phi}'(y,t) &=& -\frac{1}{2} d^2 e^{2 \sqrt{2} b \hat{\phi}} 
H^{-2/3b^2} \int^y_0 dy H^{2/3 b^2} \left( 
\frac{d V}{d \phi} + \sqrt{2} b V \right). 
\end{eqnarray}

\subsection{Derivation of effective Friedmann equation}

Here we will demonstrate how we can derive the Friedmann equation 
using the above solutions. We use the solutions which we have derived 
in the previous subsection and the junction conditions given in 
Appendix \ref{APP1}.  

Projecting $\alpha_{(1)}^{\prime}$ onto $y=0$, we get 
\begin{equation} 
\alpha_{(1)}^{\prime}(0,t) = \alpha_{(1)}^{(1)} 
=\frac{\sqrt{2}}{3}\sigma d (1-6b^2)f_{\alpha}(t)+g_{\alpha}(t)
+k_{\alpha}^{\prime}(0,t) 
- \frac{\sqrt{2}}{3}\sigma d (3b^2 -2)K_{\alpha}(t) 
+P^{\prime}(0,t). 
\end{equation}
On the other hand, the junction conditions (\ref{eq:junc_0_a}) gives  
\begin{equation}
\alpha_{(1)}^{(1)} = 
-\frac{\sqrt{2}}{6}\sigma d 
(K_{\gamma}(t)-\sqrt{2}b K_{\phi}(t) + f_{\gamma}(t) -\sqrt{2}f_{\phi}(t)) 
+P^{\prime}(0,t) 
-\frac{\kappa^2}{6}(\rho + U )d e^{\sqrt{2}b \hat{\phi}}. 
\end{equation}
So we notice  
\begin{multline}
\frac{\sqrt{2}}{3}\sigma d (1-6b^2) f_{\alpha}(t) 
+g_{\alpha}(t) 
+\frac{\sqrt{2}}{6}(f_{\gamma}(t)-\sqrt{2}b f_{\phi}(t)) 
+\frac{\kappa^2}{6}(\rho + U )d e^{\sqrt{2}b \hat{\phi}} \\
=-\frac{\sqrt{2}}{6}\sigma d 
[-2(3b^2 -2) K_{\alpha}(t)
+K_{\gamma}(t)-\sqrt{2}b K_{\phi}(t)]. 
\label{eq:Friedmann_0}
\end{multline}
There appears the homogeneous solution in eq.(\ref{eq:Friedmann_0}), 
so we have to know the explicit form of it to obtain 
the effective Friedmann equation. In order to do so, we use the junction 
condition at $y=r$.  Using it, we can write down the homogeneous 
solution in terms of the quantities on the negative tension brane. 
Indeed, projecting $\alpha_{(1)}^{\prime}$ onto $y=r$ we find 
\begin{multline}
\alpha_{(1)}^{\prime}(r,t) = \tilde{\alpha}_{(1)}^{(1)} \\
=\left[
\frac{\sqrt{2}}{3}\sigma d (1-6b^2)f_{\alpha}(t)+g_{\alpha}(t)
\right] \tilde{H}^{1-(1/3b^2)} 
+\left[
\frac{\sqrt{2}}{3}\sigma d (1-3b^2)g_{\alpha}(t)+2h_{\alpha}(t)
\right] \tilde{H}^{-1/3b^2} r \\
+\frac{\sqrt{2}}{3}\sigma d h_{\alpha}(t) 
\tilde{H}^{-1-(1/3b^2)} r^2 
+k_{\alpha}^{\prime}(r,t) 
-\frac{\sqrt{2}}{3}\sigma d (3b^2 -2)K_{\alpha}(t)  \tilde{H}^{-2/3b^2} 
+P^{\prime}(r,t),  
\end{multline} 
and the junction condition (\ref{eq:junc_r_a}) gives 
\begin{multline}
\tilde{\alpha}_{(1)}^{(1)} = 
-\frac{\sqrt{2}}{6}
[
(K_{\gamma}(t)-\sqrt{2}b K_{\phi}(t))\tilde{H}^{-2/3b^2} 
+(f_{\gamma}(t)-\sqrt{2}b f_{\phi}(t))\tilde{H}^{1-(1/3b^2)} \\
+(g_{\gamma}(t)-\sqrt{2}b g_{\phi}(t))\tilde{H}^{-1/3b^2} r 
+(h_{\gamma}(t)-\sqrt{2}b h_{\phi}(t))\tilde{H}^{-1-(1/3b^2)} r^2   
] \\
+P^{\prime}(r,t) 
+\frac{\kappa^2}{6}(\tilde{\rho}+\tilde{U})d e^{\sqrt{2}b \hat{\phi}}. 
\end{multline} 
Combining them each other, we obtain 
\begin{multline}
\left[
\frac{\sqrt{2}}{3}\sigma d (1-6b^2)f_{\alpha}(t) +g_{\alpha}(t)
+\frac{\sqrt{2}}{6}\sigma d (f_{\gamma}(t)-\sqrt{2}b f_{\phi}(t))
\right]\tilde{H}^{1-(1/3b^2)} \\
+\left[
\frac{\sqrt{2}}{3}\sigma d (1-3b^2)g_{\alpha}(t) + 2h_{\alpha}(t)
+\frac{\sqrt{2}}{6}\sigma d (g_{\gamma}(t)-\sqrt{2}b g_{\phi}(t))
\right]\tilde{H}^{-1/3b^2} r \mbox{ }\hspace{2cm} \\ 
+\left[
\frac{\sqrt{2}}{3}\sigma d h_{\alpha}(t)
+\frac{\sqrt{2}}{6}\sigma d (h_{\gamma}(t)-\sqrt{2}b h_{\phi}(t))
\right]\tilde{H}^{-1-(1/3b^2)} r^2  \\
+k_{\alpha}^{\prime}(r,t) 
-\frac{\kappa^2}{6}(\tilde{\rho}+\tilde{U})d e^{\sqrt{2}b\hat{\phi}} \\
=-\frac{\sqrt{2}}{6}\sigma d
[
-2(3b^2 -2)K_{\alpha}(t)+K_{\gamma}(t)-\sqrt{2}b K_{\phi}(t)
]\tilde{H}^{-2/3b^2}. 
\label{eq:Friedmann_r}
\end{multline}
Explicit form of all homogeneous solutions are given in Appendix
\ref{APP2}.  
Now let us use the homogeneous solution (\ref{eq:Friedmann_r}).  
Substituting (\ref{eq:Friedmann_r}) into 
(\ref{eq:Friedmann_0}), we finally obtain the effective Friedmann
equation. The concrete form of the effective Friedmann equation 
(\ref{eq:1st-order (tt)2}) is given in the next section. 
Similarly, we can derive other effective equations. All of them 
are also given in the next section.  

Here we note that one can not determine the homogeneous solution in 
one brane model because one knows the only one junction condition at 
the brane where one lives.  

\section{4D Effective equations}
\subsection{Friedmann equation, equations of motion for radion and 
scalar field}

Now we can explicitly write down four-dimensional effective equations. 
First, the evolution equations for scale factor are given by
\begin{equation} 
\ddot{\hat{\alpha}}+2\dot{\hat{\alpha}}^2 
+\frac{1}{6}\dot{\hat{\phi}}^2
= \frac{\sqrt{2}}{36} \kappa^2 \sigma 
((1-3w)\rho+4 U+3 \sqrt{2} b U')e^{-\sqrt{2}b \hat{\phi}} -
\frac{1}{6} V(\phi^{(0)}), 
\label{eq:Friedmann1} 
\end{equation} 
\begin{multline}
\ddot{\hat{\alpha}}+2\dot{\hat{\alpha}}^2 
+\frac{1}{6}\dot{\hat{\phi}}^2  
= -\frac{\sqrt{2}}{36}\kappa^2 \sigma
((1-3\tilde{w})\tilde{\rho}+4 \tilde{U}+3 \sqrt{2} b \tilde{U}')
e^{-\sqrt{2}b\phi}
\tilde{H}^{-\frac{3\Delta +4}{3\Delta +8}}  \\
-\frac{1}{6}\tilde{H}^{\frac{4}{8+3 \Delta}}V(\tilde{\phi}^{(0)})  
 + \frac{2}{3\Delta +8}
\left(
\ddot{d} +3 \dot{d}\dot{\hat{\alpha}}
+\sqrt{2}b\dot{d}\dot{\hat{\phi}}
\right)
\tilde{H}^{-1}
+\frac{3\Delta +4}{(3\Delta +8)^2}\dot{d}^2\tilde{H}^{-2}, 
\label{eq:Friedmann2}
\end{multline}
\begin{multline}
\dot{\hat{\alpha}}^2+\sqrt{2} b \dot{\hat{\alpha}}\dot{\hat{\phi}}
-\frac{1}{6}\dot{\hat{\phi}}^2  
= \frac{\sqrt{2}(\Delta + 4)}{24(1-\tilde{H}^{\zeta})} \kappa^2 \sigma
\left[
\rho+U + (\tilde{\rho} + \tilde{U} )\tilde{H}^{\frac{8}{3\Delta +8}}
\right]
e^{-\sqrt{2}b\hat{\phi}}  \\
-\frac{\sqrt{2}}{24}(\Delta+4)\sigma d 
\frac{1}{1-\tilde{H}^{\zeta}} \int^{r}_0 dy H^{\frac{8}{8+3 \Delta}}
V \\
+\frac{3\tilde{H}^{\zeta}(\Delta +4)}{1-\tilde{H}^{\zeta}}
\left[
-\frac{1}{3\Delta +8}\dot{d}\dot{\hat{\alpha}}\tilde{H}^{-1}
+\frac{1}{(3\Delta +8)^2}\dot{d}^2 \tilde{H}^{-2}
\right].
\label{eq:1st-order (tt)2}
\end{multline}
The equation for $\hat{\phi}$ is given by 
\begin{multline}
\ddot{\hat{\phi}}+3\dot{\hat{\alpha}}\dot{\hat{\phi}}
+\sqrt{2}b\dot{\hat{\phi}}^2  
= \frac{b\kappa^2 \sigma}{6(1-\tilde{H}^{\zeta})} 
\Big[
(1-3w)\rho+4 U- \sqrt{2} b^{-1}U' 
+ ((1-3\tilde{w})\tilde{\rho}+4 \tilde{U}-\sqrt{2} b^{-1} \tilde{U}')
\tilde{H}^{\frac{8}{3\Delta +8}}
\Big]
e^{-\sqrt{2}b\hat{\phi}}  \\
+\frac{\sqrt{2}}{8}(\Delta+4) \sigma d
\frac{1}{1-\tilde{H}^{\zeta}} \int^r_0 dy H^{\frac{8}{8+3 \Delta}}
\left(V'+\sqrt{2} b V \right) \\
- \frac{1}{1-\tilde{H}^{\zeta}}
\frac{\sqrt{2}b}{2} V(\phi^{(0)})\left(1- \tilde{H}^{\frac{3 \Delta+16}{3 \Delta +8}} 
\frac{V(\tilde{\phi}^{(0)})}{V(\phi^{(0)})} \right)\\
+\frac{\tilde{H}^{\zeta}}{1-\tilde{H}^{\zeta}}
\left[
-\frac{6\sqrt{2}b}{3\Delta +8}
(\ddot{d}+3\dot{d}\dot{\hat{\alpha}})\tilde{H}^{-1}
 -\frac{6\Delta +20}{3\Delta +8}
  \dot{d}\dot{\hat{\phi}}\tilde{H}^{-1}
 +\frac{24\sqrt{2}b}{(3\Delta +8)^2} \dot{d}^2 \tilde{H}^{-2}
\right].
\label{eq:scalar} 
\end{multline}
Finally the equation for $d(t)$ is given by  
\begin{multline}
\ddot{d}+3\dot{d}\dot{\hat{\alpha}}+\sqrt{2}b\dot{d}\dot{\hat{\phi}}
+\frac{3\Delta +4}{2(3\Delta +8)}\dot{d}^2 \tilde{H}^{-1}  \\
=\frac{\sqrt{2}(3\Delta +8)}{72}\tilde{H} \kappa^2 \sigma
\left[
(1-3w)\rho +4U+3 \sqrt{2}b U'
+((1-3\tilde{w})\tilde{\rho}+4 \tilde{U}+3 \sqrt{2} b \tilde{U}')
\tilde{H}^{-\frac{3\Delta +4}{3\Delta +8}}
\right]e^{-\sqrt{2}b \hat{\phi}}\\
-\frac{8+3 \Delta}{12} \tilde{H}V(\phi^{(0)}) \left(1- 
\tilde{H}^{\frac{4}{8+3 \Delta}} 
\frac{V(\tilde{\phi}^{(0)})}{V(\phi^{(0)})} \right)
, 
\label{eq:rad}
\end{multline}
where we defined 
\begin{eqnarray}
\tilde{H} &\equiv& H(r,t) = 1-d(t), \\
\zeta &\equiv& \frac{3(\Delta + 4)}{3\Delta +8}.
\end{eqnarray}
To make the spacetime non-singular, we set the negative tension 
brane at $r=1/(\sqrt{2} b^2 \sigma)$ and assume $d < 1$. 
We can also get the energy conservation law 
as follows, 
\begin{eqnarray}
\dot{\rho} + 3(1+w)\rho\dot{\hat{\alpha}}&=& 0,  \\
\dot{\tilde{\rho}}+3(1+\tilde{w}) \tilde{\rho}
\left( \dot{\hat{\alpha}}- \frac{\sqrt{2}}{3\Delta +8} \dot{d} 
\tilde{H}^{-1} \right)
&=& 0,
\end{eqnarray}
which can be derived from above equations. 
Because we have five equations for
five unknown functions $(\hat{\alpha}, \hat{\phi}, d,\rho,\tilde{\rho})$
we can consistently solve these equations.

In the RS model, the effective equations are given by
\begin{eqnarray}
\ddot{\hat{\alpha}}+2 \dot{\hat{\alpha}}^2 + \frac{1}{6} \dot{\hat{\phi}}^2
&=& \frac{\kappa^2 k}{6}((1-3w)\rho +4U)- \frac{V}{6}, \\
\ddot{\hat{\alpha}}+2 \dot{\hat{\alpha}}^2 + \frac{1}{6} \dot{\hat{\phi}}^2
&=& -\frac{\kappa^2 k}{6}((1-3 \tilde{w})\tilde{\rho} +4 \tilde{U}) 
+\ddot{d}+3\dot{\hat{\alpha}}\dot{d}-\dot{d}^2
-e^{-2 d}\frac{V}{6}, \\
\dot{\hat{\alpha}}^2-\frac{1}{6} \dot{\hat{\phi}}^2 &=& \frac{\kappa^2 k}{3}
\frac{e^{2d}}{e^{2d}-1} \left[\rho+U +(\tilde{\rho}+\tilde{U})e^{-4d} \right]
+\frac{1}{12}V(1+e^{-2d}) \nonumber \\
& & \hspace{7cm}- \frac{1}{e^{2d}-1}(2 \dot{\hat{\alpha}}\dot{d}-\dot{d}^2),
\end{eqnarray}
\begin{eqnarray}
\ddot{\hat{\phi}}+3 \dot{\hat{\alpha}} \dot{\hat{\phi}} &=&
-\kappa^2 k \frac{e^{2d}}{e^{2d}-1} (U'+\tilde{U}'e^{-4d})
-\frac{1}{4}(1+e^{-2d})V'- 2 \frac{1}{e^{2d}-1} \dot{d} \dot{\hat{\phi}},
\\
\ddot{d}+3 \dot{\hat{\alpha}} \dot{d}-\dot{d}^2
&=&\frac{\kappa^2 k}{6} \left[( (1-3w) \rho +4 U+ 
((1-3 \tilde{w}) \tilde{\rho} +4 \tilde{U}) e^{-2 d}
\right]- \frac{V}{6}(1-e^{-2d}).
\end{eqnarray}
Here we take $r=1/k$.

\subsection{Effective Theory}
In the previous section, we derived effective four-dimensional equations.
We try to interpret them in terms of the scalar-tensor gravity. 
In this section the effective theory for 
observers confined on a positive tension brane is considered. 
In this section, we do not consider the potentials $U,\ \tilde{U}$ and $V$.

The action for the scalar-tensor theory is given by 
\begin{equation}
S = \frac{1}{2\kappa_4^2}
\int d^4 x \sqrt{-g}
\left[
\varphi R -\frac{\omega(\varphi)}{\varphi}
         (\partial \varphi)^2 + {\cal L}_m
\right]. 
\label{eq:ST_action}
\end{equation}
The Friedmann equation and the equation for $\varphi$ are given by
\begin{eqnarray}
\dot{\hat{\alpha}}^2+\dot{\hat{\alpha}} \frac{\dot{\varphi}}{\varphi}
-\frac{\omega(\varphi)}{6} \left(\frac{\dot{\varphi}}{\varphi}
\right)^2 
&=& \frac{\kappa_4^2}{3} \frac{\rho}{\varphi}, \nonumber\\
\ddot{\varphi}+\dot{\varphi} \left(3 \dot{\hat{\alpha}}+ 
\frac{\dot{\omega}}{2 \omega+3} \right)
&=& \frac{\kappa_4^2}{2 \omega+3}
(1-3w) \rho.
\label{eq:BDfriedmann}
\end{eqnarray}

\subsubsection{Case1: $d = const.$}
At first, let us consider the case where the radion is somehow 
stabilized, that is, $\dot{d}=0$. In such a case, we find that eq.
(\ref{eq:1st-order (tt)2}) becomes
\begin{eqnarray}
\dot{\hat{\alpha}}^2+\sqrt{2}b\dot{\hat{\alpha}}\dot{\hat{\phi}}
-\frac{1}{6}\dot{\hat{\phi}}^2  
&=& \frac{\sqrt{2}(\Delta +4)}{24(1-\tilde{H}^{\zeta})}
e^{-\sqrt{2}b\hat{\phi}}
\left[
\rho + \tilde{\rho}\tilde{H}^{\frac{8}{3\Delta +8}}
\right], \\
\ddot{\hat{\phi}}+3\dot{\hat{\alpha}}\dot{\hat{\phi}}
+\sqrt{2}b\dot{\hat{\phi}}^2  
&=& \frac{b\kappa^2 \sigma}{6(1-\tilde{H}^{\zeta})} 
\Big[
(1-3w)\rho
+ (1-3\tilde{w})\tilde{\rho} \tilde{H}^{\frac{8}{3\Delta +8}}
\Big]e^{-\sqrt{2}b\hat{\phi}}. 
\label{eq:F1}
\end{eqnarray}
If we take
\begin{equation}
\varphi \equiv e^{\sqrt{2}b\hat{\phi}}, \quad 
\omega = \frac{6}{3\Delta +8}= \frac{1}{2b^2} , 
\label{eq:BD1}
\end{equation}
the equations become the same as (\ref{eq:BDfriedmann}).
Then the effective theory can be identified with Brans-Dicke theory with 
Brans-Dicke parameter $\omega$ given in  (\ref{eq:BD1}). 
We also find the four-dimensional gravitational constant becomes
\begin{equation}
\kappa_4^2 \equiv \frac{\sqrt{2}(\Delta +4)}{8(1-\tilde{H}^{\zeta})}
\kappa^2 \sigma, \quad 
\tilde{\kappa}_4 \equiv 
\kappa_4^2 \tilde{H}^{\frac{8}{3 \Delta+8}}.
\end{equation}
Here $\tilde{H}$ is constant because $d$ is constant now. 
It is interesting that the matter on the negative tension brane
couples to the gravity with a different gravitational constant
from the matter on a positive tension brane.

From the observational constraint, we need $\omega > 3000$ 
at least at the late stage of cosmology \cite{Will}. So the dilaton
coupling is strongly constraint as $b^2 < 1.6 \times 10^{-4}$. 
Thus in the HW theory with $b=1$, the stabilization mechanism for 
$\phi$ should be introduced in order to obtain an acceptable
late time cosmology.
 
In RS model, the theory becomes Einstein theory and the four-dimensional
gravitational constant is given by a well known formula;
\begin{equation}
\kappa_4^2= \kappa^2 k \frac{e^{2d}}{e^{2d}-1}, \quad \tilde{\kappa}_4^2
=\kappa_4^2 e^{-4d}. 
\end{equation}

\subsubsection{Case2: $\hat{\phi} = const.$}
Next, we consider the case where the bulk scalar field is stabilized, i.e., 
$\dot{\hat{\phi}}=0$.  This case has been studied by many authors 
in the context of the Randall-Sundrum model \cite{Kanno1},\cite{Kanno2},
\cite{GT}. 
In such a case, eq.(\ref{eq:1st-order (tt)2}) becomes
\begin{eqnarray}
\dot{\hat{\alpha}}^2 
-3\frac{\tilde{H}^{\zeta}(\Delta +4)}{1-\tilde{H}^{\zeta}}
\left[
-\frac{1}{3\Delta +8}\dot{\hat{\alpha}}\dot{d}\tilde{H}^{-1}
+\frac{1}{(3\Delta +8)^2}\dot{d}^2 \tilde{H}^{-2}
\right]  
= \frac{\sqrt{2}(\Delta +4)}{24(1-\tilde{H}^{\zeta})}
e^{-\sqrt{2}b\hat{\phi}}
\left[
\rho + \tilde{\rho}\tilde{H}^{\frac{8}{3\Delta +8}}
\right],
\label{eq:scalar_const_1} \\
\ddot{d}+3\dot{d}\dot{\hat{\alpha}}+\sqrt{2}b\dot{d}\dot{\hat{\phi}}
+\frac{3\Delta +4}{2(3\Delta +8)}\dot{d}^2 \tilde{H}^{-1}  
=\frac{\sqrt{2}(3\Delta +8)}{72}\tilde{H} \kappa^2 \sigma
\left[
(1-3w)\rho +(1-3\tilde{w})\tilde{\rho}
\tilde{H}^{-\frac{3\Delta +4}{3\Delta +8}}
\right]e^{-\sqrt{2}b \hat{\phi}}.
\label{eq:scalar_const}
\end{eqnarray}
From these equations, if we take
\begin{equation}
\varphi \equiv 1-\tilde{H}^{\zeta}, \quad
\omega = \frac{2}{\Delta +4} \left( \frac{\varphi}{1-\varphi} \right)
=\frac{3}{2(3 b^2+1)} \left( \frac{\varphi}{1 -\varphi} \right),
\end{equation}
eqs.(\ref{eq:scalar_const_1}) and (\ref{eq:scalar_const}) can be 
written as (\ref{eq:BDfriedmann}).
Now $\omega$ is a dynamical variable, so we conclude that scalar-tensor 
gravity is realized. 
The effective four-dimensional gravitational constant becomes 
\begin{equation}
\kappa_4^2 \equiv \frac{\sqrt{2}(\Delta +4)}{8}\kappa^2 \sigma,\quad 
\tilde{\kappa}_4^2 
\equiv \kappa_4^2(1-\varphi)^{\frac{8}{3(\Delta+4)}}=
\kappa_4^2 (1-\varphi)^{\frac{2}{3 b^2+1}}.
\end{equation}
From the observational constraint $\omega > 3000$, we find 
\begin{equation}
d > 1 - \left(\frac{1}{1500(\Delta +4)+1} \right)^{1/\zeta}. 
\end{equation}
The physical distance $D$ between the positive and negative tension brane 
is given by, 
\begin{equation}
D = \int_0^{\frac{1}{\sqrt{2}b^2 \sigma}} dy \ d \ e^{\sqrt{2}b\hat{\phi}} 
= \frac{d}{\sqrt{2}b^2 \sigma} e^{\sqrt{2}b\hat{\phi}} .
\end{equation}
Thus the constraint means the negative tension brane should be
sufficiently far away from positive tension brane. For $b \neq 0$,
the position of the singularity corresponds to $d=1$. Thus the 
constraint requires that the negative tension brane is close to 
the singularity.

In RS model with $\Delta=-8/3$, the BD field and BD parameter are 
given by
\begin{equation}
\varphi \equiv 1-e^{-2d},\quad \omega=\frac{3}{2} 
\left( \frac{\varphi}{1-\varphi} \right)
\end{equation}
and the four dimensional gravitational constant is defined by 
\begin{equation}
\kappa_4^2= \kappa^2 k,\quad \tilde{\kappa}_4^2=\kappa_4^2 (1-\varphi)^2.
\end{equation}
These results completely agree with the results obtained in 
\cite{Kanno1},\cite{Kanno2},\cite{GT}. The observational constraint implies
\begin{equation}
D= d /k > 4 /k. 
\end{equation}

\subsubsection{General Case}
In general cases, the effective theory becomes bi-scalar tensor
theory. For small $b$ satisfying the constraint $b^2 < 1.6 \times 10^{-4}$, 
we can obtain a phenomenologically acceptable cosmology without introducing a
stabilization mechanism if the negative tension brane is sufficiently
away from the positive tension brane. However, the dynamics of the moduli 
fields strongly depends on the matter contents on both branes and the initial 
conditions for the moduli fields. Thus in many cases, it seems to be inevitable
to introduce some stabilization mechanism for moduli fields. The 
stabilization problem of moduli fields are very general and serious 
problem which appears in all higher dimensional theories. Detailed 
analysis of the general dynamics of the moduli and investigation of the 
stabilization problem is beyond the scope of this paper. We will come
back to this issue in future publications\cite{KK2}.

\section{Boundary inflation}
In this section, we consider a simple model for inflating branes 
driven by potentials $U,\ \tilde{U}$ and $V$. There are 
essentially two types of the inflation \cite{Lukas1}. 
First, the potential energy 
of the brane potential $U$ can drive the inflation on our brane. 
This can be called "matter field inflation".
Second, the potential energy in the bulk can also drive 
the inflation on our brane. This can be called "modular inflation".
In general, one could have a mixture of both types of the inflation. 
In the following, we will provide a simple model for "matter field
inflation" and "modular inflation" respectively. 
For simplicity, we do not consider the matter on both branes, that 
is $\rho =\tilde{\rho}=0$ through this section.

\subsection{Matter field inflation}
First, we consider a simple model for inflation 
driven by brane potential $U$. Then we assume that
the bulk potential vanishes $V=0$. Furthermore, in order to
avoid the negative tension brane hits the positive tension
brane or singularity, we assume the moduli fields $d(t)$ and 
$\phi(t)$ are stabilized by tuning potentials $U$ and $\tilde{U}$. 
From the equations of motion for $d(t)$ and $\phi(t)$, 
we notice that this can be achieved by choosing 
\begin{equation}
U=e^{2 \sqrt{2} b \phi} U_0, \quad 
\tilde{U}=-U \tilde{H}^{\frac{3 \Delta+4}{3 \Delta+8}},
\end{equation}
where $U_0$ is independent of $\phi$. Then 
it is possible to stabilize the moduli fields $d(t)$ and $\phi(t)$;
\begin{equation}
d(t)=d_{\ast}=const. \quad \phi(t)=\phi_{\ast}=const.
\end{equation}
The effective equations become a simple equation
\begin{equation}
\dot{\hat{\alpha}}^2= \frac{\sqrt{2} (\Delta+4)}{24}e^{\sqrt{2} b \phi_{\ast}}
 \kappa^2 \sigma U_0.
\end{equation}
If one considers that the potential energy $U_0$ is provided by some
matter field on the brane, the conventional four-dimensional inflationary 
scenario can be realized. We comment on the stability of the moduli 
fields $d(t)$ and $\phi(t)$. Let us consider small fluctuations of the 
moduli fields;
\begin{equation}
d(t)=d_{\ast}+\delta d(t),\quad \phi(t)=\phi_{\ast}+\delta \phi(t).
\end{equation}
Linearizing the equations of motion for $\phi$(t) and $d(t)$, we get the 
equations for fluctuations;
\begin{eqnarray}
\ddot{\delta d}+3 \dot{\hat{\alpha}} \dot{\delta d}
+(3\Delta+4) \dot{\hat{\alpha}}^2 
\delta d=0, \\
\ddot{\delta \phi}+3 \dot{\hat{\alpha}} \dot{\delta \phi}=
-\frac{6 \sqrt{2}}{3 \Delta+8} 
\frac{\tilde{H}^{\zeta}}{1-\tilde{H}^{\zeta}} 
(\ddot{\delta d}+3 \dot{\hat{\alpha}} \dot{\delta d}).
\end{eqnarray}
Then we found that for $3 \Delta +4 >0$, the fluctuations acquire
a positive mass squared, thus the solutions are stable. On the 
other hand for $3 \Delta +4 <0$, the system of inflating two branes 
is unstable. This agrees with the result that in RS model with 
$3 \Delta +4=-4$, the inflating two branes are unstable and 
the  radion fluctuation $\delta d(t)$ has a mass squared 
$-4 \dot{\hat{\alpha}}^2$. It would be interesting to quantify 
the effect of these moduli fluctuations on the inflationary 
scenario. 

In RS model with no bulk scalar field, it is possible 
to construct a viable model without the stabilization of the 
moduli field $d(t)$. Let us consider the matter 
field inflation driven by potentials on a positive tension brane $U$.
The effective equations are given by
\begin{eqnarray}
\dot{\hat{\alpha}}^2+\frac{1}{e^{2d}-1}(2 \dot{\hat{\alpha}}\dot{d}-\dot{d}^2)
&=& \frac{\kappa^2 k}{3} \frac{e^{2d}}{e^{2d}-1}U, \\
\ddot{d}+3 \dot{\hat{\alpha}} \dot{d}-\dot{d}^2 &=& \frac{2 \kappa^2 k}{3}U.
\end{eqnarray}
The solution for $U=const.$ can be easily found as
\begin{equation}
\dot{\hat{\alpha}}^2 = \dot{d}^2 = \frac{\kappa^2 k}{3} U.
\end{equation}
The fate of branes depend on initial conditions.
If the initial condition for $d$ is properly chosen,
the inflation on positive tension brane pushes the 
negative tension brane away. Thus the observational 
constraint on $d$ becomes easy to be satisfied. 
It is also interesting to consider a model where branes 
will collide, although there are many open questions 
in cosmology based on colliding branes.  

\subsection{Modular inflation}
Next let us consider an inflation driven by bulk potential
$V$. In this case, the moduli $\phi$ acts as an inflaton. 
Thus we should not stabilize $\phi$. The 
stabilization mechanism should be switched on after inflation. 
In general, the other moduli $d(t)$ also participates in the 
dynamics of inflation. In order to simplify the model, we take 
a bulk potential as
\begin{equation}
V=- \Lambda e^{-2 \sqrt{2} b \phi} \sigma^2.
\end{equation}
Furthermore, as is done in the previous subsection, we 
stabilize $d(t)$ by tuning the brane potential $\tilde{U}$
on the negative tension brane as 
\begin{equation}
\tilde{U}= \frac{2 \sqrt{2} \sigma \Lambda}{\kappa^2 \Delta} 
(\tilde{H}^{\frac{3 \Delta+4}{3 \Delta+8}}-\tilde{H}^{-1})
e^{- \sqrt{2} b \hat{\phi}}.
\end{equation}
Then we can have $d(t)=d_{\ast}=const.$ and the effective equations
become
\begin{eqnarray}
\dot{\hat{\alpha}}^2+\sqrt{2} b \dot{\hat{\alpha}}\dot{\hat{\phi}}
-\frac{1}{6}\dot{\hat{\phi}}^2 &=& - \frac{\Delta+4}{6 \Delta} 
 \sigma \Lambda e^{-2 \sqrt{2} b \hat{\phi}}, \\
\ddot{\hat{\phi}}+3\dot{\hat{\alpha}}\dot{\hat{\phi}}
+\sqrt{2}b\dot{\hat{\phi}}^2  &=& 
-\frac{2 \sqrt{2} b}{\Delta} \sigma \Lambda
e^{-2 \sqrt{2} b \hat{\phi}}. 
\end{eqnarray}
This equations yield a power-law solutions for scale factor $e^{\hat{\alpha}}$ 
and scalar field $e^{\hat{\phi}}$.  An interesting property of these
solutions is that the power does not
depend on the amplitude of the potential $\Lambda$ and it is determined 
only by the dilaton coupling $b$. We find that a power-law inflation 
can be realized for $-2 > \Delta $.

Finally, let us consider a RS model. In this model, we can simply take 
$d \to \infty$ to neglect the contribution of the moduli $d(t)$ rather
than stabilizing $d(t)$. Then the effective equations become
\begin{eqnarray}
\dot{\hat{\alpha}}^2 &=& \frac{1}{6} \dot{\hat{\phi}}^2 + \frac{1}{12}V,
\nonumber\\
\ddot{\hat{\phi}} &+& 3 \dot{\hat{\alpha}} \dot{\hat{\phi}}+ \frac{1}{4} V'
=0. 
\end{eqnarray}
If we define $\phi_4= \hat{\phi}/\kappa k^{1/2}$ and $V_4=V/4 \kappa^2 k$,
these are completely the same as the equations obtained in 
conventional four-dimensional theory. Thus the bulk inflaton can mimic 
the four-dimensional inflaton dynamics \cite{HTS}.

\section{Conclusions and Future Works}

In this paper, we considered two branes model with bulk scalar 
field, which is a generalization of both Randall-Sundrum model and 
Ho\v{r}ava-Witten model. The bulk potential of the scalar field
has a exponential functions $V_{bulk}=\exp(-2 \sqrt{2} b \phi)$.
We considered matter on both 
branes and arbitrary potentials in the bulk and on the brane. These matter and 
potentials induce the cosmological expansion of the brane as well as the 
time evolution of the bulk scalar field and radion.
Because it is expected that the tension of the brane is much larger than 
the energy density of matter at late time cosmology, 
we have used the low-energy approximation to derive the equations 
which describe the cosmological evolution on the brane. 

The main results of this paper are the four-dimensional 
effective equations (\ref{eq:Friedmann1})-(\ref{eq:rad})
which govern the low energy dynamics of the 
two branes system. The important point is that once the 
four dimensional dynamics is determined by these effective equations, 
we can construct a correspondent full five-dimensional geometry. 
This correspondence would be useful to investigate the effect of the bulk geometry 
on four dimensional brane dynamics.

We also investigated observational constraints imposed on the model.
We assume we are living on positive tension brane.
First, we studied the case where 
the radion $d(t)$ is stabilized. In such a case, we can regard this model 
as Brans-Dicke theory with a Brans-Dicke parameter $1/2b^2$. 
From the observational  constraint, we concluded $b^2$ must be smaller than 
$1.6 \times 10^{-4}$. 
Second, we consider the case where $\hat{\phi}=const.$ We found that this model 
can be identified with the scalar-tensor theory where the Brans-Dicke 
parameter is given by $(3/2(3b^2+1)) (\varphi/1-\varphi)$, where $\varphi$
is the BD field which is defined by $d(t)$. For RS model with $b=0$, 
this agrees with the result obtained in 
\cite{Kanno1},\cite{Kanno2},\cite{GT}. 
The observational constraint implies the negative tension brane should be 
sufficiently away from the positive tension brane. In general cases
where there in no stabilization mechanism, the effective theory becomes 
bi-scalar tensor theory. 

Using the effective equations, we constructed several
simple models for inflation. The inflation can be driven by  
potentials on the brane and bulk. We can construct various 
models for boundary inflation. It is very interesting to know 
we can distinguish these models from conventional
four-dimensional model. The detailed analysis of these
inflation models including the primordial fluctuations 
will be given in the near future. 

{\bf Note added}\\
After almost completing this work, we found \cite{BBDR2}. In
\cite{BBDR2}, the equations for the radion and the bulk scalar 
are derived using a moduli approximation and qualitatively the 
same result is obtained.


\section*{Acknowledgements}
The work of K.K. was supported by JSPS.

\appendix
\begin{flushleft}
{\Large\bfseries Appendix}
\end{flushleft}

\section{Five-dimensional equations \label{APP1} }

\subsection{Einstein equations and scalar field equation}
Five-dimensional Einstein equation and scalar field equations
are given by\\
\noindent
$(y, y) : $
\begin{multline}
\alpha^{\prime 2}+\alpha^{\prime}\beta^{\prime}
          -(\ddot{\alpha}+2\dot{\alpha}^2
          -\dot{\alpha}\dot{\beta})e^{2(\gamma-\beta)} \\
=\frac{1}{6}\phi^{\prime 2}
            +\frac{1}{6}\dot{\phi}^2 e^{2(\gamma-\beta)}
-\frac{1}{6}
\left[
\left(b^2-\frac{2}{3}\right)
                 e^{2(\gamma-\sqrt{2}b\phi)}\sigma^2
-V(\phi) e^{2\gamma}
\right].
\label{eq:(yy)-component}
\end{multline}
$(t, t) : $
\begin{multline}
\alpha^{\prime\prime}+2\alpha^{\prime 2}
                   -\gamma^{\prime}\alpha^{\prime}
          -(\dot{\alpha}^2+\dot{\gamma}\dot{\alpha})e^{2(\gamma-\beta)} \\
=-\frac{1}{6}\phi^{\prime 2}
            -\frac{1}{6}\dot{\phi}^2 e^{2(\gamma-\beta)}
-\frac{1}{6}
\left[
\left(b^2-\frac{2}{3}\right)
                 e^{2(\gamma-\sqrt{2}b\phi)}\sigma^2
-V(\phi) e^{2\gamma}
\right] \\
+\frac{1}{3}e^{\gamma}\big\{ [-\sqrt{2}e^{-\sqrt{2}b\phi}\sigma
                      -\kappa^2 (\rho+U)] \delta(y)
+[\sqrt{2}e^{-\sqrt{2}b\phi}\sigma
     -\kappa^2 (\tilde{\rho}+\tilde{U}) ]\delta(y-r) \big\}.
\end{multline}
$(i, j) : $
\begin{multline}
2\alpha^{\prime\prime}-2\gamma^{\prime}\alpha^{\prime}
           +2\alpha^{\prime}\beta^{\prime}+3\alpha^{\prime 2}
           +\beta^{\prime\prime}+\beta^{\prime 2}
            -\beta^{\prime}\gamma^{\prime} \\
-(2\ddot{\alpha}+2\dot{\gamma}\dot{\alpha}
                  -2\dot{\alpha}\dot{\beta}+3\dot{\alpha}^2
                  +\ddot{\gamma}+\dot{\gamma}^2
   -\dot{\beta}\dot{\gamma}) e^{2(\gamma-\beta)} \hspace{5cm} \\
=-\frac{1}{2}\phi^{\prime 2}
            +\frac{1}{2}\dot{\phi}^2 e^{2(\gamma-\beta)}
            -\frac{1}{2}
\left[
\left(b^2-\frac{2}{3}\right)
                 e^{2(\gamma-\sqrt{2}b\phi)}\sigma^2
-V(\phi) e^{2\gamma}
\right] \\
+ e^{\gamma}\big\{ [-\sqrt{2}e^{-\sqrt{2}b\phi}\sigma
                      +\kappa^2 (p-U) ]\delta(y) 
+[ \sqrt{2}e^{-\sqrt{2}b\phi}\sigma
    +\kappa^2 (\tilde{p}-\tilde{U}) ] \delta(y-r) \big\}.
\end{multline}
$(y, t) : $
\begin{equation}
\dot{\alpha}^{\prime}-\dot{\gamma}\alpha^{\prime}
          -\dot{\alpha}(\beta^{\prime}-\alpha^{\prime})
        =-\frac{1}{3}\dot{\phi}\phi^{\prime}.
\end{equation}
Equation of motion for $\phi : $
\begin{multline}
\phi^{\prime\prime}+3\alpha^{\prime}\phi^{\prime}
          +\beta^{\prime}\phi^{\prime}
           -\gamma^{\prime}\phi^{\prime}
-(\ddot{\phi}+3\dot{\alpha}\dot{\phi}
-\dot{\beta}\dot{\phi}+\dot{\gamma}\dot{\phi})e^{2(\gamma-\beta)} \\
=-\sqrt{2}b\left(b^2-\frac{2}{3}\right)
                          e^{2(\gamma-\sqrt{2}b\phi)}\sigma^2 
+\frac{1}{2}\frac{dV}{d\phi} e^{2\gamma} \hspace{6cm}\\
+e^{\gamma}\left[
 \left(-2be^{-\sqrt{2}b\phi}\sigma
            +\kappa^2 \frac{d U}{d \phi}
 \right)\delta(y)
+\left(2be^{-\sqrt{2}b\phi}\sigma
               +\kappa^2 
           \frac{d \tilde{U}}{d \phi} 
 \right)\delta(y-r)\right].
\label{eq:phi-component}
\end{multline}
We expand these equations in terms of $\varepsilon$.

\subsection{0th-oder}
The field equations of the 0-th order are given as follows, 
\begin{equation}
\alpha_{(0)}^{\prime 2}+\alpha_{(0)}^{\prime}\beta_{(0)}^{\prime}
=\frac{1}{6}\phi_{(0)}^{\prime 2}
-\frac{1}{6}\left(b^2-\frac{2}{3}\right)\sigma^2
 e^{2(\gamma_{(0)}-\sqrt{2}b\phi_{(0)})}, 
\end{equation}
\begin{equation}
\alpha_{(0)}^{\prime\prime}+2\alpha_{(0)}^{\prime 2}
-\gamma_{(0)}^{\prime}\alpha_{(0)}^{\prime}
=-\frac{1}{6}\phi_{(0)}^{\prime 2}
-\frac{1}{6}\left(b^2-\frac{2}{3}\right)\sigma^2
e^{2(\gamma_{(0)}-\sqrt{2}b\phi_{(0)})}, 
\end{equation}
\begin{multline}
2\alpha_{(0)}^{\prime\prime}-2\gamma_{(0)}^{\prime}\alpha_{(0)}^{\prime}
+2\alpha_{(0)}^{\prime}\beta_{(0)}^{\prime}+3\alpha_{(0)}^{\prime 2}
+\beta_{(0)}^{\prime\prime}+\beta_{(0)}^{\prime 2}
-\beta_{(0)}^{\prime}\gamma_{(0)}^{\prime} \\
=-\frac{1}{2}\phi_{(0)}^{\prime 2}
-\frac{1}{2}\left(b^2-\frac{2}{3}\right)\sigma^2
e^{2(\gamma_{(0)}-\sqrt{2}b\phi_{(0)})},
\end{multline}
\begin{equation}
\dot{\alpha}_{(0)}^{\prime}-\dot{\gamma}_{(0)}\alpha_{(0)}^{\prime}
-\dot{\alpha}_{(0)}(\beta_{(0)}^{\prime}-\alpha_{(0)}^{\prime})
=-\frac{1}{3}\dot{\phi}_{(0)}\phi_{(0)}^{\prime},
\end{equation}
\begin{equation}
\phi_{(0)}^{\prime\prime}+3\alpha_{(0)}^{\prime}\phi_{(0)}^{\prime}
+\beta_{(0)}^{\prime}\phi_{(0)}^{\prime}
-\gamma_{(0)}^{\prime}\phi_{(0)}^{\prime}
=-\sqrt{2}b\left(b^2-\frac{2}{3}\right)\sigma^2
e^{2(\gamma_{(0)}-\sqrt{2}b\phi_{(0)})}.
\end{equation}

Next, we can get the junction conditions at $y=0$ as follows,
\begin{eqnarray}
\alpha_{(0)}^{(1)}(t) &=& -\frac{\sqrt{2}}{6}\sigma
e^{\gamma_{(0)}-\sqrt{2}b\phi_{(0)}}|_{y=0}, \\
\beta_{(0)}^{(1)}(t) &=& -\frac{\sqrt{2}}{6}\sigma
e^{\gamma_{(0)}-\sqrt{2}b\phi_{(0)}}|_{y=0}, \\
\phi_{(0)}^{(1)}(t) &=& 
-b\sigma \ e^{\gamma_{(0)}-\sqrt{2}b\phi_{(0)}}|_{y=0}.
\end{eqnarray}
Similarly, we can get the junction conditions at $y=r$ as follows,
\begin{eqnarray}
\tilde{\alpha}_{(0)}^{(1)}(t) &=& -\frac{\sqrt{2}}{6}\sigma
e^{\gamma_{(0)}-\sqrt{2}b\phi_{(0)}}|_{y=r}, \\
\tilde{\beta}_{(0)}^{(1)}(t) &=& -\frac{\sqrt{2}}{6}\sigma
e^{\gamma_{(0)}-\sqrt{2}b\phi_{(0)}}|_{y=r}, \\
\tilde{\phi}_{(0)}^{(1)}(t) &=& 
-b\sigma \ e^{\gamma_{(0)}-\sqrt{2}b\phi_{(0)}}|_{y=r}.
\end{eqnarray}

\subsection{1st order}

In this subsection, we consider the equations of the order 
$\varepsilon^1$, 
which determine the behavior of $\hat{\alpha}(t), \hat{\phi}(t)$ 
and $d(t)$. 
They are given as, 
\begin{multline}
2\alpha_{(0)}^{\prime}\alpha_{(1)}^{\prime}
+\alpha_{(0)}^{\prime}\beta_{(1)}^{\prime}
+\alpha_{(1)}^{\prime}\beta_{(0)}^{\prime}
-\left[
\ddot{\alpha}_{(0)}+2\dot{\alpha}_{(0)}^2
-\dot{\alpha}_{(0)}\dot{\beta}_{(0)}\right]e^{2(\gamma_{(0)}-\beta_{(0)})}\\
=\frac{1}{3}\phi_{(0)}^{\prime}\phi_{(1)}^{\prime}
+\frac{1}{6}\dot{\phi}_{(0)}^2 e^{2(\gamma_{(0)}-\beta_{(0)})}
-\frac{1}{3}\left(b^2-\frac{2}{3}\right) \sigma^2
\left(\gamma_{(1)}-\sqrt{2}b\phi_{(1)}\right)
e^{2(\gamma{(0)}-\sqrt{2}b\phi{(0)})}+\frac{1}{6}e^{2 \gamma_{(0)}} V,
\label{eq:1st-order (yy)}
\end{multline}
\begin{multline}
\alpha_{(1)}^{\prime\prime}+4\alpha_{(0)}^{\prime}\alpha_{(1)}^{\prime}
-\gamma_{(0)}^{\prime} \alpha_{(1)}^{\prime}
-\gamma_{(1)}^{\prime} \alpha_{(0)}^{\prime}
-\left(\dot{\alpha}_{(0)}^2
+\dot{\gamma}_{(0)}\dot{\alpha}_{(0)}\right) 
e^{2(\gamma_{(0)}-\beta_{(0)})} \\
=-\frac{1}{3}\phi_{(0)}^{\prime}\phi_{(1)}^{\prime}
-\frac{1}{6}\dot{\phi}_{(0)}^2 e^{2(\gamma_{(0)}-\beta_{(0)})}
-\frac{1}{3}\left(b^2-\frac{2}{3}\right) \sigma^2
\left(\gamma_{(1)}-\sqrt{2}b\phi_{(1)}\right)
e^{2(\gamma_{(0)}-\sqrt{2}b\phi_{(0)})} 
+\frac{1}{6}e^{2 \gamma_{(0)}} V, \\
\label{eq:1st-order (tt)}
\end{multline}
\begin{multline}
2\alpha_{(1)}^{\prime\prime}-2\gamma_{(0)}^{\prime}\alpha_{(1)}^{\prime}
-2\gamma_{(1)}^{\prime}\alpha_{(0)}^{\prime}
+2\alpha_{(0)}^{\prime}\beta_{(1)}^{\prime}
+2\alpha_{(1)}^{\prime}\beta_{(0)}^{\prime}
+6\alpha_{(0)}^{\prime}\alpha_{(1)}^{\prime}
+\beta_{(1)}^{\prime\prime}
+2\beta_{(0)}^{\prime}\beta_{(1)}^{\prime}
-\beta_{(0)}^{\prime}\gamma_{(1)}^{\prime}
-\beta_{(1)}^{\prime}\gamma_{(0)}^{\prime} \\
+\left(
-2\ddot{\alpha}_{(0)}-2\dot{\gamma}_{(0)}\dot{\alpha}_{(0)}
+2\dot{\alpha}_{(0)}\dot{\beta}_{(0)}
-3\dot{\alpha}_{(0)}^2
-\ddot{\gamma}_{(0)}-\dot{\gamma}_{(0)}^2
+\dot{\beta}_{(0)}\dot{\gamma}_{(0)}
\right)e^{2(\gamma_{(0)}-\beta_{(0)})} \\
=-\phi_{(0)}^{\prime}\phi_{(1)}^{\prime}
+\frac{1}{2}\dot{\phi}_{(0)}^2 e^{2(\gamma_{(0)}-\beta_{(0)})}
-\left(b^2-\frac{2}{3}\right)\sigma^2
\left(\gamma_{(1)}-\sqrt{2}b\phi_{(1)}\right)
e^{2(\gamma_{(0)}-\sqrt{2}b\phi_{(0)})}+\frac{1}{2} 
e^{2 \gamma_{(0)}} V, 
\label{eq:1st-order (ij)}
\end{multline}
\begin{multline}
\dot{\alpha}_{(1)}^{\prime}
-(\dot{\gamma}_{(0)}\alpha_{(1)}^{\prime}
 +\dot{\gamma}_{(1)}\alpha_{(0)}^{\prime}) \\
-\left[\dot{\alpha}_{(0)}(\beta_{(1)}^{\prime}-\alpha_{(1)}^{\prime})
+\dot{\alpha}_{(1)}(\beta_{(0)}^{\prime}-\alpha_{(0)}^{\prime})\right] 
=-\frac{1}{3}\left(
\dot{\phi}_{(0)}\phi_{(1)}^{\prime}
+\dot{\phi}_{(1)}\phi_{(0)}^{\prime}\right),
\label{eq:1st-order (yt)}
\end{multline}
\begin{multline}
\phi_{(1)}^{\prime\prime}
+3\alpha_{(0)}^{\prime}\phi_{(1)}+3\alpha_{(1)}^{\prime}\phi_{(0)}^{\prime}
+\beta_{(0)}^{\prime}\phi_{(1)}^{\prime}
+\beta_{(1)}^{\prime}\phi_{(0)}^{\prime}
-\gamma_{(0)}^{\prime}\phi_{(1)}^{\prime}
-\gamma_{(1)}^{\prime}\phi_{(0)}^{\prime} \\
-\left(
\ddot{\phi}_{(0)}+3\dot{\alpha}_{(0)}\dot{\phi}_{(0)}
-\dot{\beta}_{(0)}\dot{\phi}_{(0)}
+\dot{\gamma}_{(0)}\dot{\phi}_{(0)} \right) 
e^{2(\gamma{(0)}-\beta_{(0)})}  \\
=-2 \sqrt{2} b\left(b^2-\frac{2}{3}\right) \sigma^2
\left(\gamma_{(1)}-\sqrt{2}b\phi_{(1)}\right)
e^{2(\gamma_{(0)}-\sqrt{2}b\phi_{(0)})} + \frac{1}{2} \frac{d V}{d \phi}.
\label{eq:1st-order phi}
\end{multline}
Similarly, we can derive junction conditions at $y=0$ and $y=r$ 
as follows, 
\begin{eqnarray}
\alpha_{(1)}^{(1)}(t) &=& -\frac{\sqrt{2}}{6}
\left(
\gamma_{(1)}-\sqrt{2}b\phi_{(1)}
\right)\sigma
e^{
\gamma_{(0)}-\sqrt{2}b\phi_{(0)} } 
-\frac{\kappa^2}{6} (\rho+U) \ e^{ \gamma_{(0)} } \ \big|_{y=0},
\label{eq:junc_0_a} \\
\beta_{(1)}^{(1)}(t) &=& -\frac{\sqrt{2}}{6}
\left(
\gamma_{(1)}-\sqrt{2}b\phi_{(1)}
\right)\sigma
e^{
\gamma_{(0)}-\sqrt{2}b\phi_{(0)}
}
+\frac{\kappa^2}{6}(3 p+2 \rho -U) \ e^{
\gamma_{(0)}
} \ \big|_{y=0}, \\
\phi_{(1)}^{(1)}(t) &=& 
-b\left(
\gamma_{(1)}-\sqrt{2}b\phi_{(1)}
\right)\sigma
e^{
\gamma_{(0)}-\sqrt{2}b\phi_{(0)}
}
+\frac{\kappa^2}{2} \frac{d U}{d \phi} \ e^{\gamma_{(0)}
}\ \big|_{y=0}, \\
\tilde{\alpha}_{(1)}^{(1)}(t) &=& -\frac{\sqrt{2}}{6}
\left(
\gamma_{(1)}-\sqrt{2}b\phi_{(1)}
\right)\sigma
e^{
\gamma_{(0)}-\sqrt{2}b\phi_{(0)}
}
+\frac{\kappa^2}{6}(\tilde{\rho}+\tilde{U}) \ e^{
\gamma_{(0)}
}\ \big|_{y=r},
\label{eq:junc_r_a} \\
\tilde{\beta}_{(1)}^{(1)}(t) &=& -\frac{\sqrt{2}}{6}
\left(
\gamma_{(1)}-\sqrt{2}b\phi_{(1)}
\right)\sigma
e^{
\gamma_{(0)}-\sqrt{2}b\phi_{(0)}
}
-\frac{\kappa^2}{6}(3 \tilde{p}+2 \tilde{\rho} - \tilde{U}) \ e^{
\gamma_{(0)}
}\ \big|_{y=r}, \\
\tilde{\phi}_{(1)}^{(1)}(t) &=& 
-b\left(
\gamma_{(1)}-\sqrt{2}b\phi_{(1)}
\right)\sigma
e^{
\gamma_{(0)}-\sqrt{2}b\phi_{(0)}
}
-\frac{\kappa^2}{2} \frac{d \tilde{U}}{d \phi}\ e^{\gamma_{(0)}
}\ \big|_{y=r}. 
\end{eqnarray}

\newpage

\section{Solutions for 1st Order Equations \label{APP2} } 

\subsection{Homogeneous solutions}

Here we will write down the homogeneous solutions in terms of 
$\tilde{\rho}, \tilde{U}$.  

\begin{eqnarray}
K_{\alpha} &=& -\frac{\sqrt{2}\kappa^2 \sigma^{-1}}{3\Delta +8}
(\tilde{\rho}+\tilde{U}) e^{\sqrt{2}b\hat{\phi}} \nonumber \\ 
& &+\frac{2\sqrt{2}\sigma^{-2}}{\Delta} e^{2\sqrt{2}b\hat{\phi}}
\Bigg[
-\frac{2\sqrt{2}}{(\Delta +4)}
\left(
\dot{\hat{\alpha}}^2
+\sqrt{2}b\dot{\hat{\alpha}}\dot{\hat{\phi}}
-\frac{1}{6}\dot{\hat{\phi}}^2
\right) \tilde{H}^{\frac{3(\Delta +4)}{3\Delta +8}} \nonumber \\
& &
+\frac{6\sqrt{2}}{3\Delta +8}
\dot{d}\dot{\hat{\alpha}} \tilde{H}^{\frac{4}{3\Delta +8}}
+\frac{6\sqrt{2}}{(3\Delta +8)^2}
\dot{d}^2  \tilde{H}^{-\frac{3\Delta +4}{3\Delta +8}} 
\Bigg] \nonumber \\
& &-\frac{\sqrt{2}\sigma^{-1}d}{3\Delta}e^{2\sqrt{2}b\hat{\phi}}
\tilde{H}^{-\frac{8}{3\Delta +8}}
\int_0^r dy H^{\frac{8}{3\Delta +8}} V,   \\ 
K_{\beta} &=& \frac{\sqrt{2}\kappa^2 \sigma^{-1}}{3\Delta +8}
((3\tilde{w}+2)\tilde{\rho}-\tilde{U}) e^{\sqrt{2}b\hat{\phi}} \nonumber \\
& &+\frac{2\sqrt{2}\sigma^{-2}}{\Delta} e^{2\sqrt{2}b\hat{\phi}}
\Bigg[
-\frac{2\sqrt{2}}{(\Delta +4)}
\left(
2\ddot{\hat{\alpha}}
+\dot{\hat{\alpha}}^2
+\sqrt{2}b\ddot{\hat{\phi}}
+\frac{3\Delta +13}{6}\dot{\hat{\phi}}^2
\right) \tilde{H}^{\frac{3(\Delta +4)}{3\Delta +8}} \nonumber \\
& &
+\frac{6\sqrt{2}}{3\Delta +8}
\left(
\frac{\ddot{d}}{d}+2\sqrt{2}b\frac{\dot{d}}{d}\dot{\hat{\phi}}
\right) \tilde{H}^{\frac{4}{3\Delta +8}}
-\frac{6\sqrt{2}}{(3\Delta +8)^2}
\dot{d}^2  \tilde{H}^{-\frac{3\Delta +4}{3\Delta +8}} 
\Bigg] \nonumber \\
& &-\frac{\sqrt{2}\sigma^{-1}d}{\Delta}e^{2\sqrt{2}b\hat{\phi}}
\tilde{H}^{-\frac{8}{3\Delta +8}}
\int_0^r dy H^{\frac{8}{3\Delta +8}} 
\left[
\sqrt{2}b V^{\prime} +(1+2b^2) V
\right]
-\frac{\sigma^{-2}}{\Delta}e^{2\sqrt{2}b\phi}H V, \\ 
K_{\phi} &=& \frac{\sqrt{2}\kappa^2 \sigma^{-1}}{\Delta}
U^{\prime}e^{\sqrt{2}b\hat{\phi}} \\
& &+\frac{2\sqrt{2}\sigma^{-2}}{\Delta} e^{2\sqrt{2}b\hat{\phi}}
\Bigg[
-\frac{2\sqrt{2}}{(\Delta +4)}
\left(
\ddot{\hat{\phi}}
+3\dot{\hat{\alpha}}\dot{\hat{\phi}}
+\sqrt{2}b\dot{\hat{\phi}}^2
-3\sqrt{2}b
\left( 
\ddot{\hat{\alpha}}
+2\dot{\hat{\alpha}}^2
+\frac{1}{6}\dot{\hat{\phi}}^2 
\right)
\right) \nonumber \\
& &+\frac{6\sqrt{2}}{3\Delta +8}
\dot{d}\dot{\hat{\phi}}\tilde{H}^{\frac{4}{3\Delta +8}}
-\frac{36b}{(3\Delta +8)^2}
\dot{d}^2  \tilde{H}^{-\frac{3\Delta +4}{3\Delta +8}} 
\Bigg] \nonumber \\
& &-\frac{\sqrt{2}\sigma^{-1}d}{\Delta}e^{2\sqrt{2}b\hat{\phi}}
\tilde{H}^{-\frac{8}{3\Delta +8}}
\int_0^r dy H^{\frac{8}{3\Delta +8}} (V+\sqrt{2}bV). 
\end{eqnarray}

\subsection{Particular solutions}

The particular solutions for 1st order equations are given in 
eqs.(\ref{eq:particular_alpha})-(\ref{eq:particular_phi}). 
The concrete form of them becomes as follows;
\begin{eqnarray}
f_{\alpha} &=& 
\frac{4\sigma^{-2}}{(\Delta +4)(\Delta +2)} e^{2\sqrt{2}b\hat{\phi}}
\left[
\dot{\hat{\alpha}}^2 +\sqrt{2}b\dot{\hat{\alpha}}\dot{\hat{\phi}}
-\frac{1}{6}\dot{\hat{{\phi}}}^2 
-\frac{3(\Delta +4)}{3\Delta +4}
 \left(
 \frac{\dot{d}}{d}\dot{\hat{\alpha}}+\frac{1}{2}\frac{\dot{d}^2}{d^2} 
 \right)
\right], \\
f_{\beta} &=& 
\frac{4\sigma^{-2}}{(\Delta +4)(\Delta +2)} e^{2\sqrt{2}b\hat{\phi}}
\Bigg[
2\ddot{\hat{\alpha}}+\dot{\hat{\alpha}}^2 
+\sqrt{2}b \ddot{\hat{\phi}}
+\frac{3\Delta +13}{6}\dot{\hat{{\phi}}}^2 \nonumber \\
& &\hspace{7cm}-\frac{3(\Delta +4)}{3\Delta +4}
 \left(
 2\sqrt{2}b\frac{\dot{d}}{d}\dot{\hat{\phi}}
 +\frac{\ddot{d}}{d}
-\frac{1}{2}\frac{\dot{d}^2}{d^2} 
 \right)
\Bigg], \\
f_{\phi} &=& 
\frac{4\sigma^{-2}}{(\Delta +4)(\Delta +2)} e^{2\sqrt{2}b\hat{\phi}}
\Bigg[
\ddot{\hat{\phi}} +3\dot{\hat{\alpha}}\dot{\hat{\phi}}
+\sqrt{2}b\dot{\hat{\phi}}^2
-3\sqrt{2}b
\left(
\ddot{\hat{\alpha}}
+2\dot{\hat{\alpha}}^2
+\frac{1}{6}\dot{\hat{\phi}}^2
\right) \nonumber \\ 
& &\hspace{8cm}-\frac{3(\Delta +4)}{3\Delta +4}
\left(
 \frac{\dot{d}}{d}\dot{\hat{\phi}}
 +\frac{3\sqrt{2}}{2}b\frac{\dot{d}^2}{d^2} 
 \right)
\Bigg],  \\
f_{\gamma} &=& \sqrt{2}bf_{\phi}, \\
g_{\alpha} &=& 
-\frac{3\sqrt{2}\sigma^{-1}}{3\Delta +4} e^{2\sqrt{2}b\hat{\phi}} 
\left(
2\dot{d}\dot{\hat{\alpha}} +\frac{\dot{d}^2}{d} 
\right),   \\
g_{\beta} &=& 
-\frac{3\sqrt{2}\sigma^{-1}}{3\Delta +4} e^{2\sqrt{2}b\hat{\phi}} 
\left(
2\ddot{d}+4\sqrt{2}b\dot{d}\dot{\hat{\phi}}
-\frac{\dot{d}^2}{d}
\right),  \\
g_{\phi} &=& 
-\frac{18\sigma^{-1}}{3\Delta +4} e^{2\sqrt{2}b\hat{\phi}}
\left(
b\frac{\dot{d}^2}{d} +\frac{\sqrt{2}}{3}\dot{d}\dot{\hat{\phi}} 
\right),  \\
g_{\gamma} &=& \sqrt{2}bg_{\phi}, \\
h_{\alpha} &=& 
-\frac{1}{4}\dot{d}^2 e^{2\sqrt{2}b\hat{\phi}}, \\
h_{\beta} &=& 
\frac{1}{4}\dot{d}^2 e^{2\sqrt{2}b\hat{\phi}}, \\
h_{\phi} &=& 
-\frac{3\sqrt{2}}{4}b\dot{d}^2 e^{2\sqrt{2}b\hat{\phi}}, \\
h_{\gamma} &=& \sqrt{2}b h_{\phi}.
\end{eqnarray}

For $\Delta = -2$ and/or $\Delta = -4/3$, the particular solutions 
should be treated separately. 
But even in such cases, we get the same effective 4D equations 
as we have shown in Sec.4.   

\section{Homogeneous and Particular Solutions in RS model \label{APP3} }

\subsection{Homogeneous solutions}

\begin{eqnarray}
\alpha_{(1)}(y,\ t) &=& K_{\alpha}(t)e^{4kdy} + F_{\alpha}(t) +P(y,\ t),  \\
\beta_{(1)}(y,\ t) &=& K_{\beta}(t)e^{4kdy} + F_{\beta}(t) 
                           +P(y,\ t),  \\
\gamma_{(1)}(y,\ t) &=& K_{\gamma}(t)e^{4kdy}
                         -\frac{3\sqrt{2}}{d}P^{\prime}(y,\ t), \\
\phi_{(1)}(y,\ t) &=& K_{\phi}(t)e^{4kdy} + F_{\phi}(t), 
\end{eqnarray} 
where $F_{\alpha}, F_{\beta}, F_{\phi}$ and $P$ are arbitrary 
functions which do not affect the resultant 4D equations. 
$K_{\alpha}, K_{\beta}, K_{\gamma}$ and $K_{\phi}$ yield the the
constraints;
\begin{equation}
3K_{\alpha}+K_{\beta}+K_{\gamma}=0, 
\end{equation}
\begin{equation}
\dot{K}_{\alpha}+(K_{\alpha}-K_{\beta})\dot{\hat{\alpha}}
+\frac{1}{3}K_{\phi}\dot{\hat{\phi}}
+\frac{1}{4}\dot{K}_{\gamma}=0.
\end{equation}
Here we define $k = (\sqrt{2}/6)\sigma $ as in Sec.2.3. 

\subsection{Particular solutions} 

The particular solutions are given by 
\begin{eqnarray}
\alpha_{(1)} &=& \left[
-\frac{1}{4k^2} \left(
 \dot{\hat{\alpha}}^2
 -\frac{1}{6}\dot{\hat{\phi}}
 +\frac{\dot{d}}{d}\dot{\hat{\alpha}}
 +\frac{1}{2}\frac{\dot{d}^2}{d^2} \right)
+\frac{1}{2k}
          \left(\dot{d}\dot{\hat{\alpha}}
                 +\frac{1}{2}\frac{\dot{d}^2}{d^2} \right) y 
-\frac{1}{4}\dot{d}^2 y^2 \right]
\ e^{2 k d y},  \\ 
\beta_{(1)} &=& \left[
-\frac{1}{4k^2} 
\left(
  2\ddot{\hat{\alpha}}
 +\dot{\hat{\alpha}}^2
 +\frac{5}{6}\dot{\hat{\phi}}
 +\frac{\ddot{d}}{d}
 -\frac{1}{2}\frac{\dot{d}^2}{d^2} \right) 
+\frac{1}{2k}
          \left(\ddot{d}
                 -\frac{1}{2}\frac{\dot{d}^2}{d^2} \right) y  
+\frac{1}{4}\dot{d}^2 y^2
\right]
\ e^{2 k d y}, \\ 
\phi_{(1)} &=& \left[
-\frac{1}{4k^2}
\left(
\ddot{\hat{\phi}}
+3\dot{\hat{\alpha}}\dot{\hat{\phi}}
+\frac{\dot{d}}{d}\dot{\hat{\phi}}
\right)
+\frac{1}{2k}\dot{d}\dot{\hat{\phi}}
\right]
\ e^{2 k d y}, \\ 
\gamma_{(1)} &=& 0. 
\end{eqnarray}



\end{document}